\documentclass[a4paper,12pt]{article}

\usepackage{epsfig}
\input{psfig}
\usepackage{amssymb}
\begin{document}

\title{\bf QCD Studies at HERA\footnote{Talk given
at the Intern. Conference on New Trends in High Energy Physics, \newline
22-29.9.2001, Yalta, Crimea, Ukraine.} \\
Selected Topics
}
\author{J. OLSSON \\ (on behalf of the H1 and ZEUS collaborations) \\
     DESY, Notkestra\ss e 85, 22603 Hamburg, Germany \\
        E-mail: jan.olsson @ desy.de}
\date{ }
\maketitle
\begin{abstract}
Several topics from the wide field of QCD studies in Deep-inelastic 
$ep$ Scattering
at HERA are addressed. They
include QCD analyses of the inclusive cross section 
with the determination of $\alpha_s$ and the proton gluon density from
the $F_2$ scaling violations, and the determination of the longitudinal
structure function $F_L$. 
QCD analyses of inclusive jet and
dijet data are also presented. 
Finally jet substructure and three-jet production are discussed. 
\end{abstract}
%
%
\section{Introduction}

Studies of Deep-inelastic lepton-nucleon scattering (DIS) have played a
fundamental role in establishing QCD as the strong interaction theory and
in exploring the parton structure of the nucleon. 
With the advent of HERA, in which  
electrons or positrons of 27.5 GeV energy collide 
with protons of 820 GeV (in the 
last years 920 GeV), the tests of QCD have been extended by 
several orders of
magnitude with respect to the range in Bjorken-$x$ and in $Q^2$,
the squared momentum transfer between lepton and nucleon.
The early fixed target experiments observed
scaling violations, i.e. the variation of the structure functions 
with $Q^2$. The scaling violations are well described 
by QCD, in which they are related to the gluon density in the proton,
and to the strong interaction coupling constant, $\alpha_s$.
\par\noindent 
At HERA the low $x$ region was experimentally
explored for the first time, and the first measurements\cite{earlyF2}
of the proton structure function $F_2(x,Q^2)$  revealed a steep
rise of $F_2$ towards low $x$ values. 
indicating a high gluon density in the proton at low $x$.
A key question is then the validity of the DGLAP evolution 
equations\cite{DGLAP} at
low $x$ values, since in the DGLAP evolution higher order terms proportional
to $\alpha_s\cdot ln(1/x)$ are neglected. 
One expects that at some value of $x$ non-linear gluon interaction effects
will become important, damping the rise of the cross section in accordance 
with unitarity requirements. While this question cannot yet be answered,
the low~$x$ region remains a key area for future QCD studies at HERA.
\vspace*{0.4cm}
\par\noindent
\begin{minipage}[t]{14cm}
\begin{minipage}[t]{4.5cm}
\begin{minipage}[t]{4.5cm}
\epsfig{file=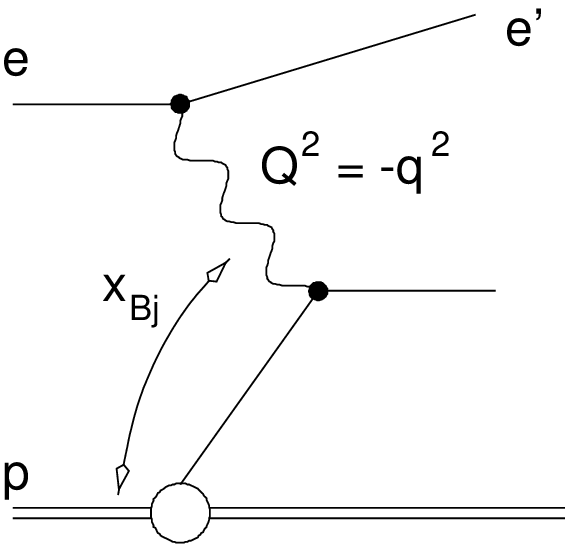,width=4.5cm}
\end{minipage}
\par\noindent
\hspace*{0.3cm}
\begin{minipage}[t]{4.1cm}
Figure 1: Diagram for lowest order $ep$ DIS 
\end{minipage}
\end{minipage}
\vspace*{-5.4cm}
\par\noindent
\hspace*{4.8cm}
\begin{minipage}[t]{9cm}
At the same time as precision measurements of the inclusive DIS cross section
and the proton structure functions
are being pursued, QCD tests of processes of higher order in 
$\alpha_s$ are also being investigated. These tests involve the study of jet
production in DIS, a field for which HERA is well suited 
 with its large $ep$ CMS energy $\sqrt{s}=300-320$ GeV.
This experimental work is intimately connected with theory 
development, since tests of 
higher order jet production need the corresponding processes 
to be quantitatively estimated.   
\end{minipage}
\end{minipage}

\section{The Inclusive DIS Cross Section and the Proton Structure}

Inclusive $ep$ neutral current (NC) 
DIS is to lowest order, ${\cal O}(\alpha_s^0)$, described
by the diagram in Fig.~1, in which a virtual boson is emitted by the 
electron and scatters off a parton in the proton. In the one-photon
exchange approximation, the double differential cross section 
can be written as 
\vspace*{0.2cm}\par\noindent
\centerline{$
{d^2 \sigma\over {dx dQ^2}} = {2 \pi \alpha^2\over 
{x Q^4}} (Y_{+} F_2(x,Q^2) - y^2 F_L(x,Q^2)), \qquad Y_{+} = 1 + (1-y)^2 .
$}
\par\noindent
The inelasticity $y=Q^2/xs$ represents the energy fraction transferred to the
 proton in the scattering process.
$F_2$ and $F_L$ are the (unpolarized, electromagnetic) 
proton structure functions and contain the information
about the momentum distribution of partons in the proton.
They are not calculable in theory 
and have to be measured. 
However, when measured at a given $Q^2$ value, their evolution in $Q^2$ can
be predicted in theory.
\par\noindent
The NC DIS cross section can also be written as
\vspace*{0.2cm}\par\noindent
\centerline{$
{d^2 \sigma\over {dx dQ^2}} = \Gamma (y) (\sigma_T + \epsilon (y)\sigma_L)
$}
\par\noindent
where the $ep$ scattering process is considered as the interaction of a flux
of virtual photons with the proton. Here
$\Gamma (y) = Y_{+} \alpha / (2 \pi Q^2 x)$ is the flux factor and 
$\epsilon (y) = 2 (1-y) / Y_+$ defines the photon polarization.
$\sigma_{T}$ and $\sigma_{L}$ are the cross sections of the interaction
of transverse and longitudinally polarized photons with the proton. 
These cross
sections are related to the structure functions: 
\vspace*{0.2cm}\par\noindent
\centerline{$
F_2(x,Q^2) = {Q^2\over {4 \pi^2 \alpha}} (\sigma_{T}(x,Q^2) + 
\sigma_{L}(x,Q^2))$ \ \ and  \ \ 
$F_L(x,Q^2) = {Q^2\over {4 \pi^2 \alpha}} \sigma_{L}(x,Q^2) .$}
\par\noindent
Due to cross section positivity,
the relation  $0 \le F_L \le F_2$ is obeyed.
\smallskip
\par\noindent
In the QPM world, without gluons, $\sigma_L = 0$, since longitudinally
polarized photons do not interact with massless spin 1/2 partons.
 Thus, in QPM also
$F_L = 0$\cite{CallanGross}.
In QCD quarks radiate gluons and interact through gluon exchange.
Radiated gluons in turn can
split into quark-antiquark pairs (``sea quarks'') or gluons. The gluon 
radiation
results in a transverse momentum component of the quarks, which can now
also couple to longitudinally polarized photons. Thus, $\sigma_{L}$ and
$F_L$ get non-zero values.  
Due to its origin, $F_L$ is directly dependent on the 
gluon distribution in the proton and therefore the measurement of $F_L$
provides a sensitive test of QCD at low $x$ values.
In fact, the low~$x$ region cannot be understood without measuring 
$F_L$ precisely.
\medskip
\par\noindent
\begin{minipage}[t]{14cm}
\begin{minipage}[t]{7cm}
\begin{minipage}[t]{7cm}
\hspace*{-0.2cm}
\epsfig{file=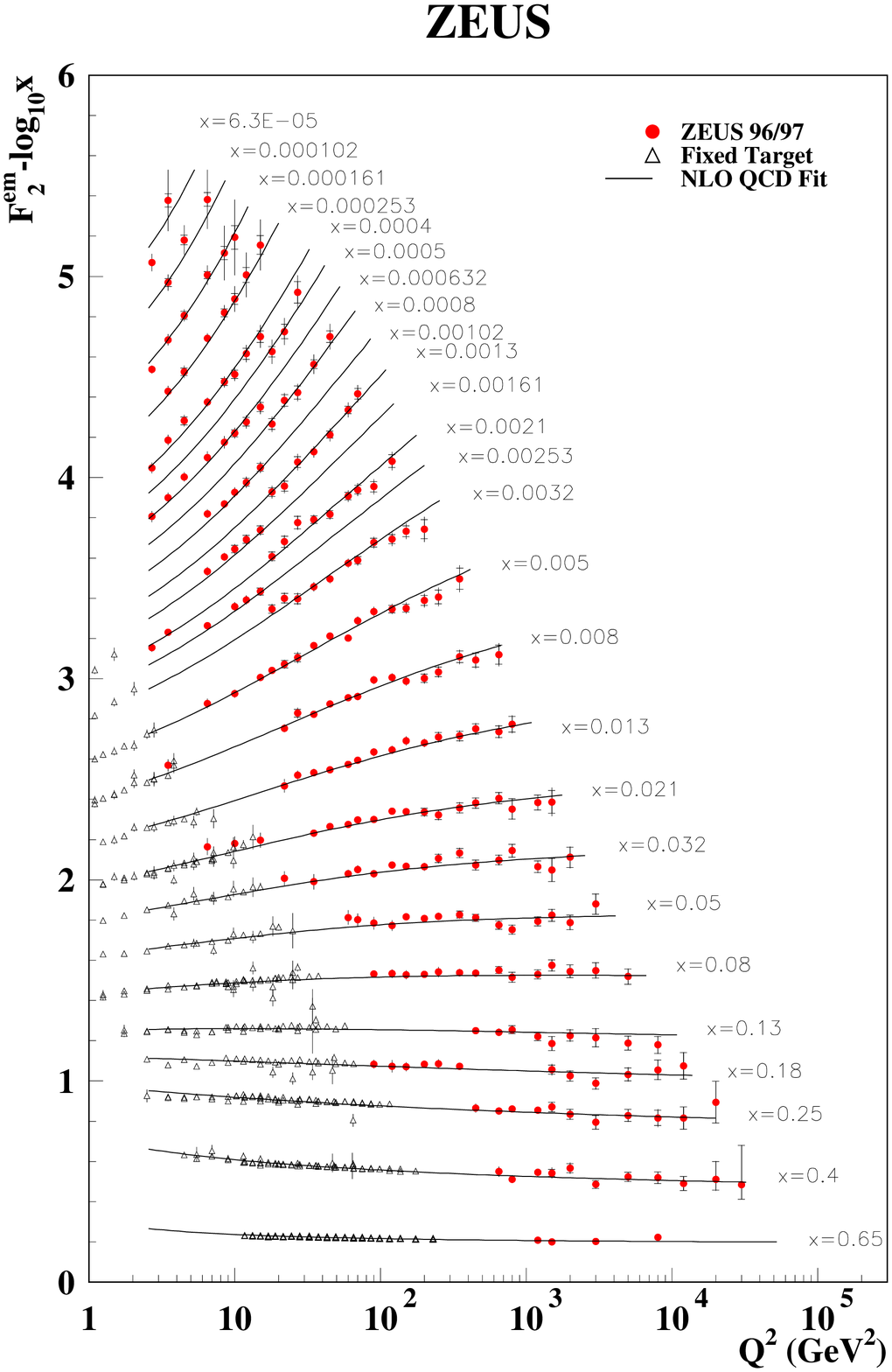,width=7cm}
\end{minipage}
\par\noindent
\hspace*{0.2cm}
\begin{minipage}[t]{6cm}
Figure 2: $F_2$ vs $Q^2$, for fixed values of $x$. Data from ZEUS and
fixed target experiments. The curves show the NLO QCD fit to the data.
\end{minipage}
\end{minipage}
\vspace*{-12.0cm}
\par\noindent
\hspace*{6.9cm}
\begin{minipage}[t]{7.0cm}
The $ep$ cross section is usually written in the ``reduced'' form, in which 
the $Q^2$ dependence due to the photon propagator is removed, \newline
\centerline{$  
\sigma_{r} \equiv F_2(x,Q^2) - {y^2\over {Y_{+}}} {F_L(x,Q^2)} . 
$}
Since the contribution of the longitudinal structure
function $F_L$ to the cross section can be sizeable only at large 
values of $y$, in a large kinematic range the relation 
$\sigma_{r} \approx F_2$ holds to a very good approximation.
\par\noindent
The H1 and ZEUS collaborations have recently 
presented high statistics measurements\cite{H199107,H100181,ZEUSF201} of the
NC DIS cross section and extracted $F_2$ in the kinematic range
$1.5-2.7<Q^2<30000$~GeV$^2$ and $3-6\cdot 10^{-5}<x<0.65$. The ZEUS data 
are shown in Fig.~2. The strong (positive) scaling violations at 
low $x$ values, due to the increase of the gluon density  
($\frac{\partial F_{2}}{\partial \log Q^{2}} 
\sim \alpha_{s} xg$ at low $x$),
are clearly seen.
At large $x$ values negative scaling violations
appear
($\frac{\partial F_{2}}{\partial \log Q^{2}} \sim \alpha_{s} F_{2}$
at large $x$).
\end{minipage}
\end{minipage}
\smallskip
\par\noindent
The ZEUS and H1 data are
in very good agreement, both with each other and,
at the largest $x$ values where data overlap,   
with the earlier fixed target experiments.
\par\noindent
Both collaborations have subjected their $F_2$ data to extensive 
QCD analyses in next-to-leading-order (NLO), 
extracting the parton density functions (PDFs) 
and $\alpha_s$\cite{H100181,ZEUSNLOfit2001}. While the general analysis and fit strategy is 
similar in both
collaborations, there are also many differences in the details, e.g.
in the density parametrizations and in the treatment of flavour number and 
in the use of the fixed target data. An extensive discussion can be found in
\cite{NaganoRingberg2001}.
\smallskip
\par\noindent
Fig.~2 shows that the NLO QCD fits, which are based
on the DGLAP evolution, describe the data
very well over no less than four
orders of magnitude and down to surprisingly low
$Q^2$ of a few GeV$^2$.
The gluon 
density extracted in the NLO QCD fits is shown for both experiments in Fig.~3
for three values of $Q^2$. Within the error bands there is reasonable
agreement.
Like $F_2$, the gluon density increases towards low $x$ values, and 
the increase is steeper with increasing $Q^2$.
\par\noindent
\begin{minipage}[t]{14cm}
\begin{minipage}[t]{7cm}
\begin{minipage}[t]{7cm}
\epsfig{file=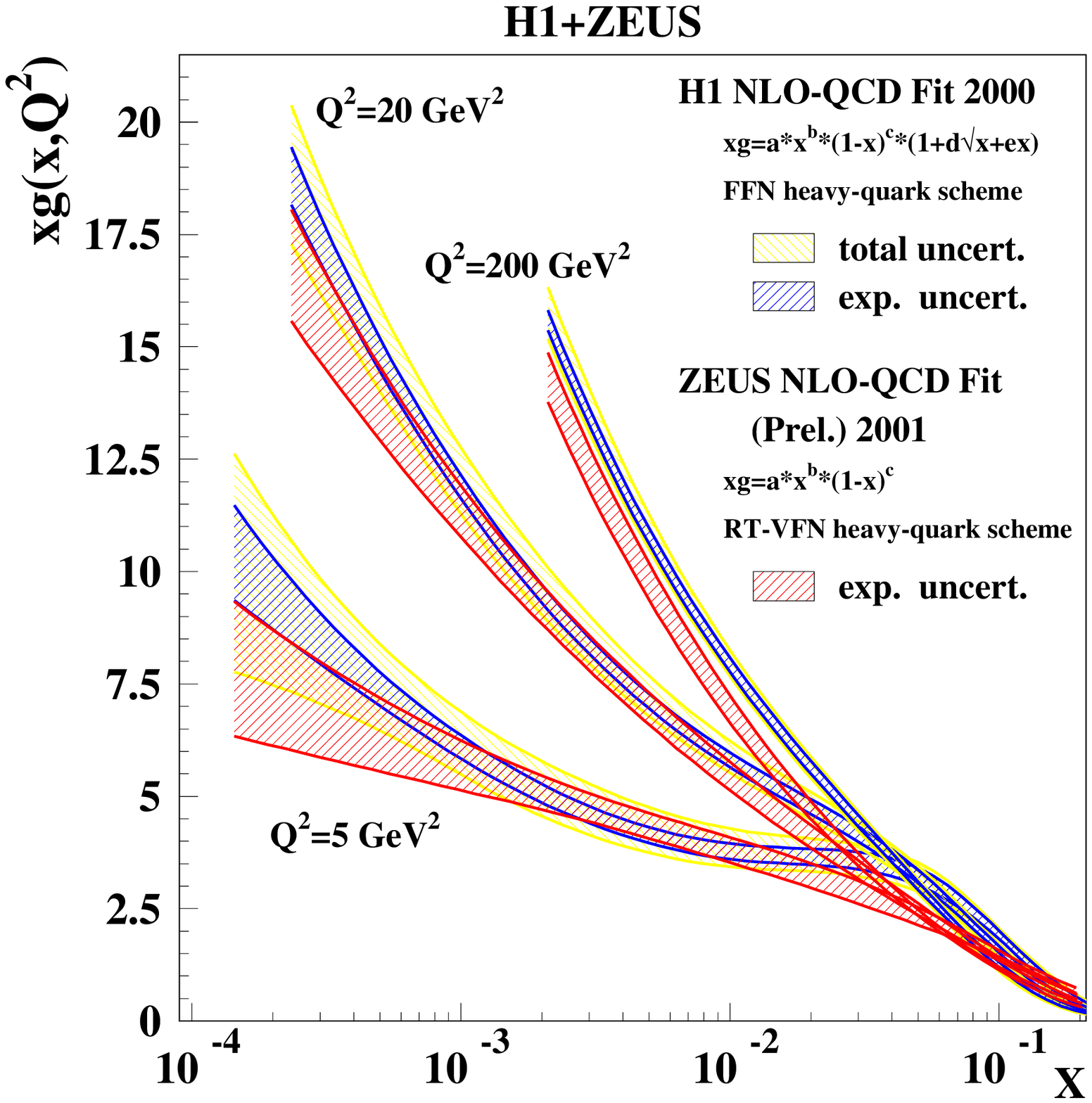,width=7cm}
\end{minipage}
\par\noindent
\hspace*{0.5cm}
\begin{minipage}[t]{6cm}
Figure 3: Gluon density for three different values of $Q^2$. The error bands
show the experimental and (for H1) also the total (including theoretical)
uncertainty.
\end{minipage}
\end{minipage}
\vspace*{-9.2cm}
\par\noindent
\hspace*{7.7cm}
\begin{minipage}[t]{6cm}
\begin{minipage}[t]{6cm}
\epsfig{file=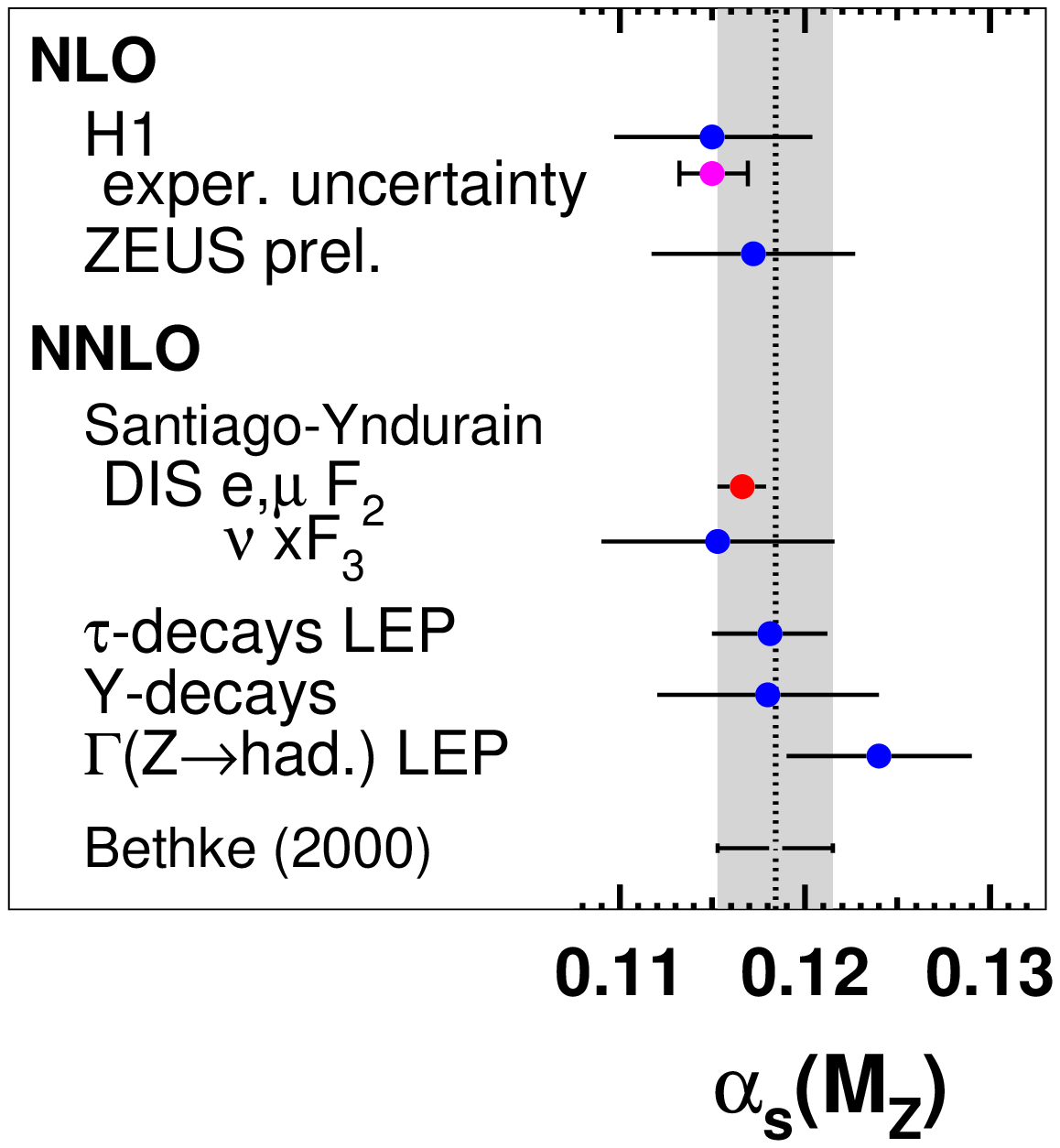,width=6cm}
\end{minipage}
\par\noindent
\begin{minipage}[t]{6cm}
Figure 4: 
$\alpha_s$ values obtained in the NLO QCD analyses of $F_2$ data. Also 
shown are $\alpha_s$ values obtained in a recent 
NNLO analysis\cite{SYNNLO} as well as other $\alpha_s$ values obtained
at NNLO level. 
\end{minipage}
\end{minipage}
\end{minipage}
\bigskip
\smallskip
\par\noindent
The $\alpha_s$ values obtained in the NLO fits of H1 and ZEUS are shown in 
Fig.~4\footnote{Figure courtesy M. Erdmann, 
LP2001, Rome 2001\cite{Erdmann2001}.}. 
In the H1 case also the error due only to the experimental uncertainty
is shown. Thus, the theoretical uncertainty dominates the total error. The
major part of this error is due to the choice of renormalization scale, and 
to a lesser extent, also to the choice of the factorization scale. This is
expected to change once next-to-next-to-leading-order (NNLO) 
QCD calculations are available. Fig.~4 also shows
the result of a recent NNLO analysis of $ep$ DIS data\cite{SYNNLO}, in 
which the moments of $F_2$ are fitted to obtain an impressingly 
small error on $\alpha_s$. Comparing the $ep$ DIS $\alpha_s$ result
 with the result obtained
when using the more imprecise $\nu N$ scattering data, 
or with the $\alpha_s$ values obtained in NNLO for
$\tau$-decays, $\Upsilon$-decays, or in analysis of $\Gamma (Z)$, it is 
obvious that the $ep$ DIS data are 
very competitive in precision determinations
of $\alpha_s$. Note that the analysis in \cite{SYNNLO} used earlier data on
$F_2$\cite{H1ZEUSF296}, and not yet the most recent, very precise data.
\par\noindent 
Both collaborations are working on further reduction of the experimental
errors on the $F_2$ data, errors which are 
completely dominated by systematics,
in particular the uncertainty on the calorimetric energy scale. 
Clearly, theoretical progress is essential in order to profit from 
future, still better experimental precision in the $F_2$ data. 

\section{The Longitudinal Structure Function \boldmath$F_L$}

A fully independent measurement of $F_L$ at HERA, in a wide range of
$x$ and $Q^2$, is only possible via a substantial variation of the beam
energies; at any given values of $x$ and $Q^2$ 
the difference of the reduced
cross section, measured at two values of the CMS energy $s=Q^2/xy$, is a 
direct measure of $F_L$:
\par\noindent
\centerline{
$
\sigma_{r,1} - \sigma_{r,2} = 
[(y^2 / Y_+)_1 - (y^2 / Y_+)_2] \cdot F_L
$.}
\medskip
\par\noindent
\begin{minipage}[t]{14cm}
\begin{minipage}[t]{5cm}
\begin{minipage}[t]{5cm}
\epsfig{file=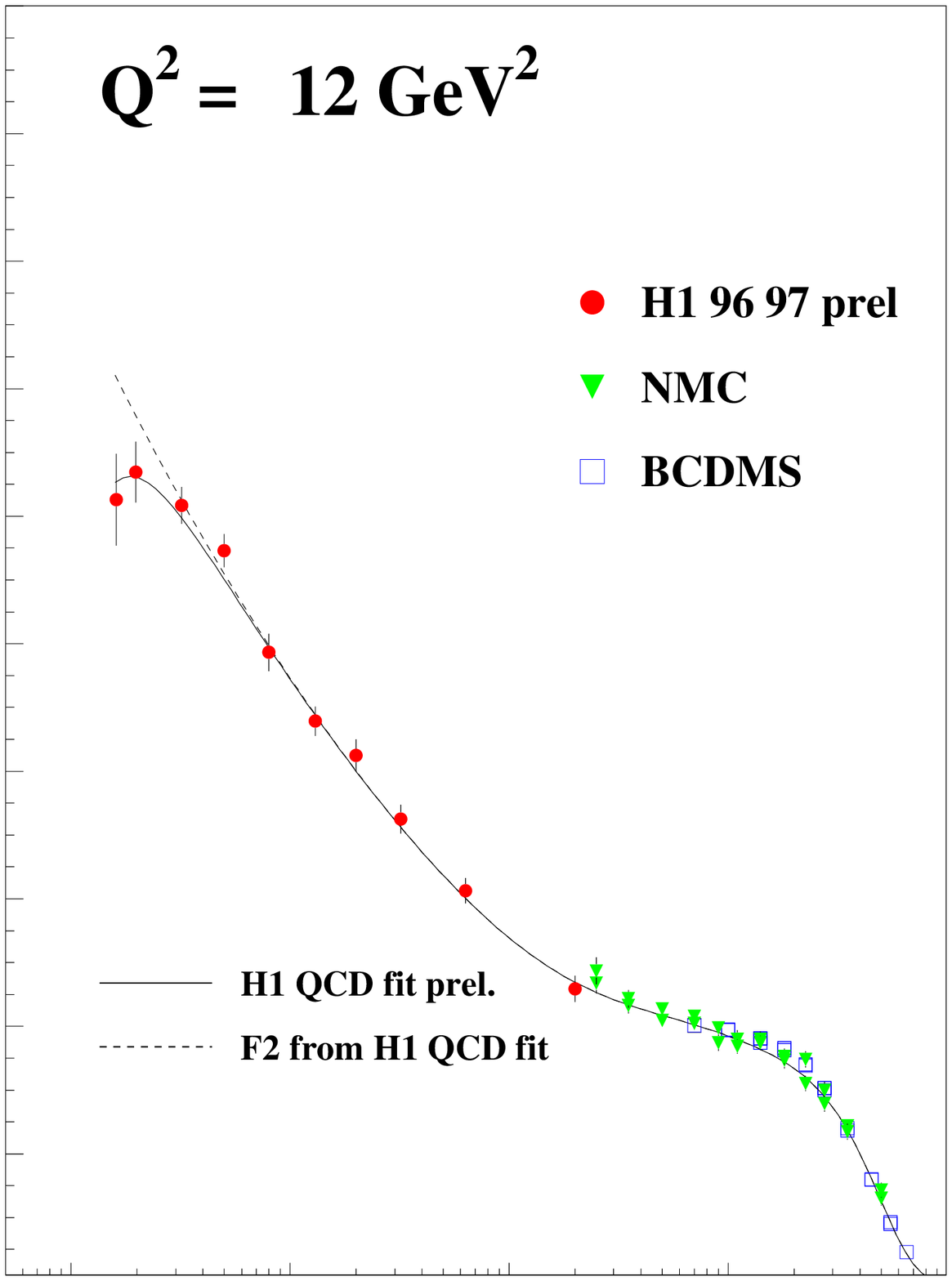,width=5cm}
\end{minipage}
\par\noindent
\hspace*{0.3cm}
\begin{minipage}[t]{4.3cm}
Figure 5: Reduced cross section vs. $x$, for $Q^2=12$ GeV$^2$. The curves
are due to the H1 NLO QCD fit and represent $F_2$ and the cross section.
\end{minipage}
\end{minipage}
\vspace*{-9.9cm}
\par\noindent
\hspace*{5.3cm}
\begin{minipage}[t]{8.6cm}
This measurement is not yet performed but is 
part of the HERA~II physics program\cite{BGK1996}.
While waiting for this to happen, the
H1 collaboration has performed several indirect extractions of $F_L$. The
methods of extraction are based on the direct measurement of the cross
section in the high $y$ region, and 
on the extrapolation of the precise knowledge of $F_2$
at low values of $y$, into the region of high $y$. Cross section measurements 
of the NC DIS process in the high $y$ region pose a challenge to the HERA
experiments, since low energy (down to 3 GeV is achieved)  scattered 
electrons have to be triggered, identified 
and well measured. The detailed understanding of the detector and of the 
photoproduction background is essential and 
an important part of the measurement
is therefore the improvement of both calorimetry  
and the detection of charged tracks in the H1 central and backward regions.
The measurement of track charge and momentum at all scattering angles is
particularly important in 
\end{minipage}
\end{minipage}
\smallskip
\par\noindent
estimating the amount and shape 
of the background from photon conversions
(with photons from the $\pi^{\circ}\to\gamma\gamma$ decay), 
through the identification of "wrong" charge 
electrons.   
\par\noindent
Two methods are used for the  extraction of $F_L$ from the cross
section measurements,
namely the "subtraction" method applied at $Q^2>10$ GeV$^2$,
and the "derivative" method, applied for $Q^2<10$ GeV$^2$. 
The subtraction method is illustrated in Fig.~5, showing the reduced cross
section as function of $x$ at fixed $Q^2=12$~GeV$^2$. At the lowest $x$
values (corresponding to high $y$ values) 
the cross section falls below the $F_2$ curve, extrapolated via the
H1 NLO QCD fit. Thus, $F_L$ is obtained from the 
difference $F_2 - \sigma_r$. This method was first explored 
in \cite{H1FL96} and in the recent analyses, using data 
from 1996-97\cite{H100181} and 1998-2000\cite{H1FLBudapest01}, $F_L$ is 
extracted up to $Q^2$ values of 700 GeV$^2$. Fig.~6 shows
the $Q^2$ dependence of $F_L$, separately for $e^+p$ and $e^-p$ data. As it
should be, $F_L$ is independent of the lepton beam. One also sees that $F_L$
is clearly different from the extreme possibilities, $F_L=0$ or 
$F_L=F_2$. Within the precent precision of the data there is good agreement
with the QCD expectation, as projected using the H1 NLO fit.
\par\noindent
In the second method of extracting $F_L$, 
the derivative of the reduced cross section with
respect to $\ln y$ is formed: 
\medskip
\par\noindent
\centerline{$
\left({\partial \sigma_{r}\over {\partial \ln y}}\right)_{Q^2} = 
\left({\partial  F_2\over {\partial \ln y}}\right)_{Q^2} - F_{L} 
       \cdot 2y^2 \cdot {{2 - y}\over {Y^2_{+}}} - {\partial F_{L}
        \over {\partial \ln y}} \cdot {y^2\over {Y_{+}}} .
$}
\vspace*{-0.8cm}
\par\noindent
\begin{minipage}[t]{14cm}
\begin{minipage}[t]{7cm}
\begin{minipage}[t]{7cm}
\epsfig{file=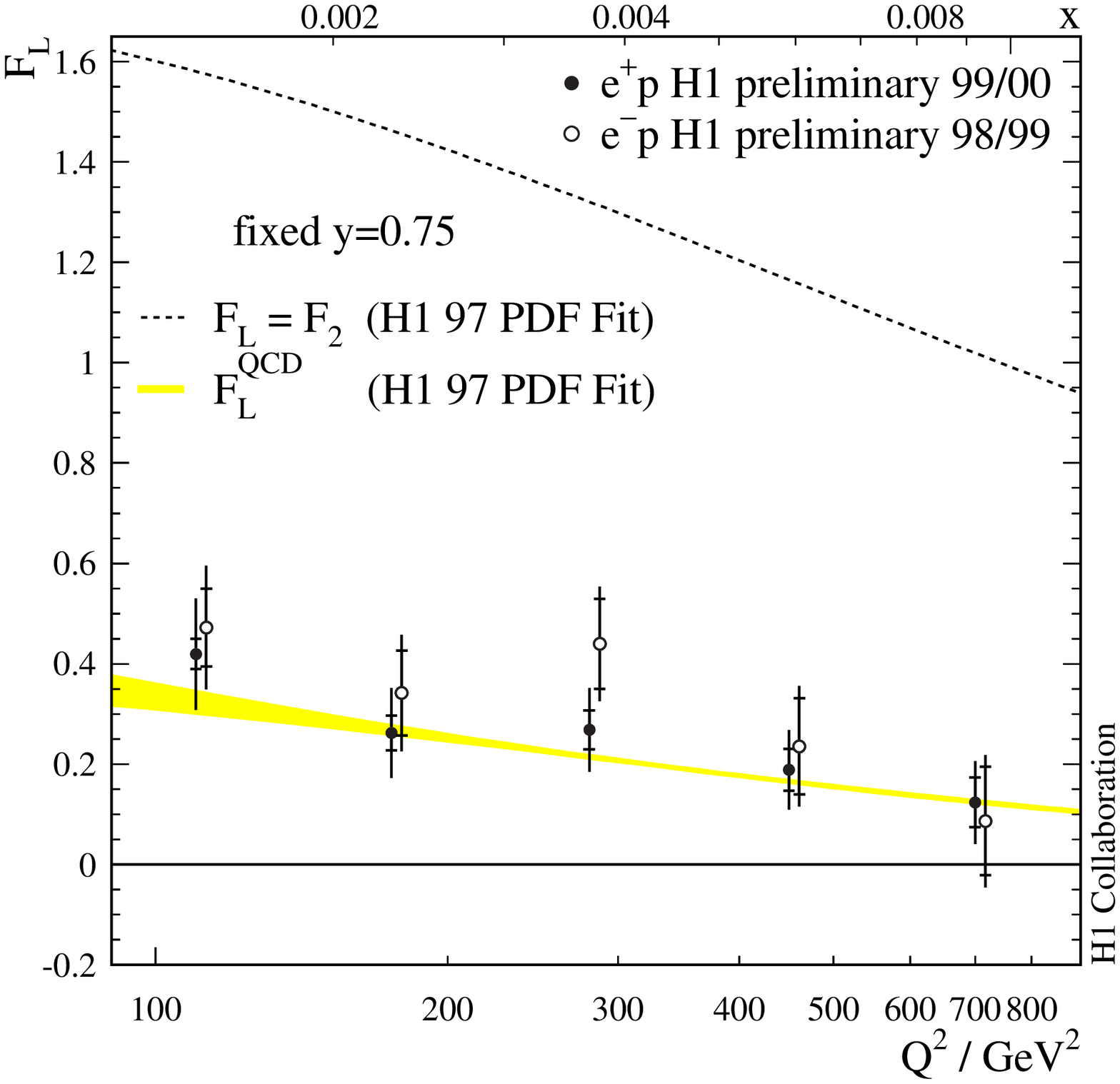,width=7cm}
\end{minipage}
\par\noindent
\hspace*{0.5cm}
\begin{minipage}[t]{6cm}
Figure 6: $F_L$ vs. $Q^2$ (or, equivalently, vs. $x$, upper scale),
for fixed $y=0.75$, for $e^+p$ and $e^-p$ data. 
The curves show the H1 NLO QCD fit, obtained for several assumptions on 
$F_L$. The shaded band shows the expectation for $F_L$ and its uncertainty
from the QCD fit.
\end{minipage}
\end{minipage}
\vspace*{-11.4cm}
\par\noindent
\hspace*{7.2cm}
\begin{minipage}[t]{6.7cm}
For $y \rightarrow 1$ the cross section derivative tends to the limit 
$(\partial F_2 / \partial \ln y)_{Q^2} - {2\cdot F_{L}}$, neglecting the
contribution from the derivative of $F_{L}$.
At largest $y$ the $F_{L}$ contribution dominates the derivative of the 
reduced cross section $\sigma_{r}$. This is in contrast to the influence of
$F_{L}$ on $\sigma_{r}$ which is dominated by the contribution of $F_2$ for 
all $y$. A further advantage of the derivative method is that it can be applied
down to very low $Q^2 \simeq$ 1 GeV$^2$ where a QCD description of $F_2(x,Q^2)$
is complicated due to higher order and possible non-perturbative corrections.
The measured cross section derivative 
is shown in Fig.~7 as a function of $y$, in several bins of $Q^2$. 
Since for a given $Q^2$ value,
 $F_2 \sim x^{- \lambda} \sim e^{\lambda \ln y}$, and since 
$\lambda$ is small for low values
of $Q^2$, 
$\frac{\partial F_{2}}{\partial \ln y}$ 
is linear in $\ln y$ to good approximation. 
This is clearly seen in Fig.~7.
In each $Q^2$ bin straight line fits were made to the de- 
\end{minipage}
\end{minipage}
\smallskip
\par\noindent
rivative data for $y <$ 0.3. The line
fits describe the data very well and
the extrapolation of the straight line was taken to represent the contribution 
of $F_2$ at high $y$. 
The small contribution of $\partial F_{L} / \partial \ln y$ to 
the derivative was corrected for by using NLO QCD and the  
correction was included in the overall error of the measured $F_{L}$.
The derivative method was used in the analysis of the low $Q^2$ data, both
in the 1996-97 data\cite{H100181} and in the 1999-2000 
data\cite{H1FLBudapest01}. Since the latter data have higher $p$ beam energy,
920 Gev as compared to 820 GeV in the earlier data, the accessible $x$ 
range could be 
extended to even smaller values. 
It is also worthwhile to note that the H1 NLO fit is based on the 1996-97 
data; when using this fit
in the analyses of the later data proper account was taken
in order to correct for the difference in $p$ beam energy.
\medskip 
\par\noindent
\begin{minipage}[t]{14cm}
\begin{minipage}[t]{9cm}
\epsfig{file=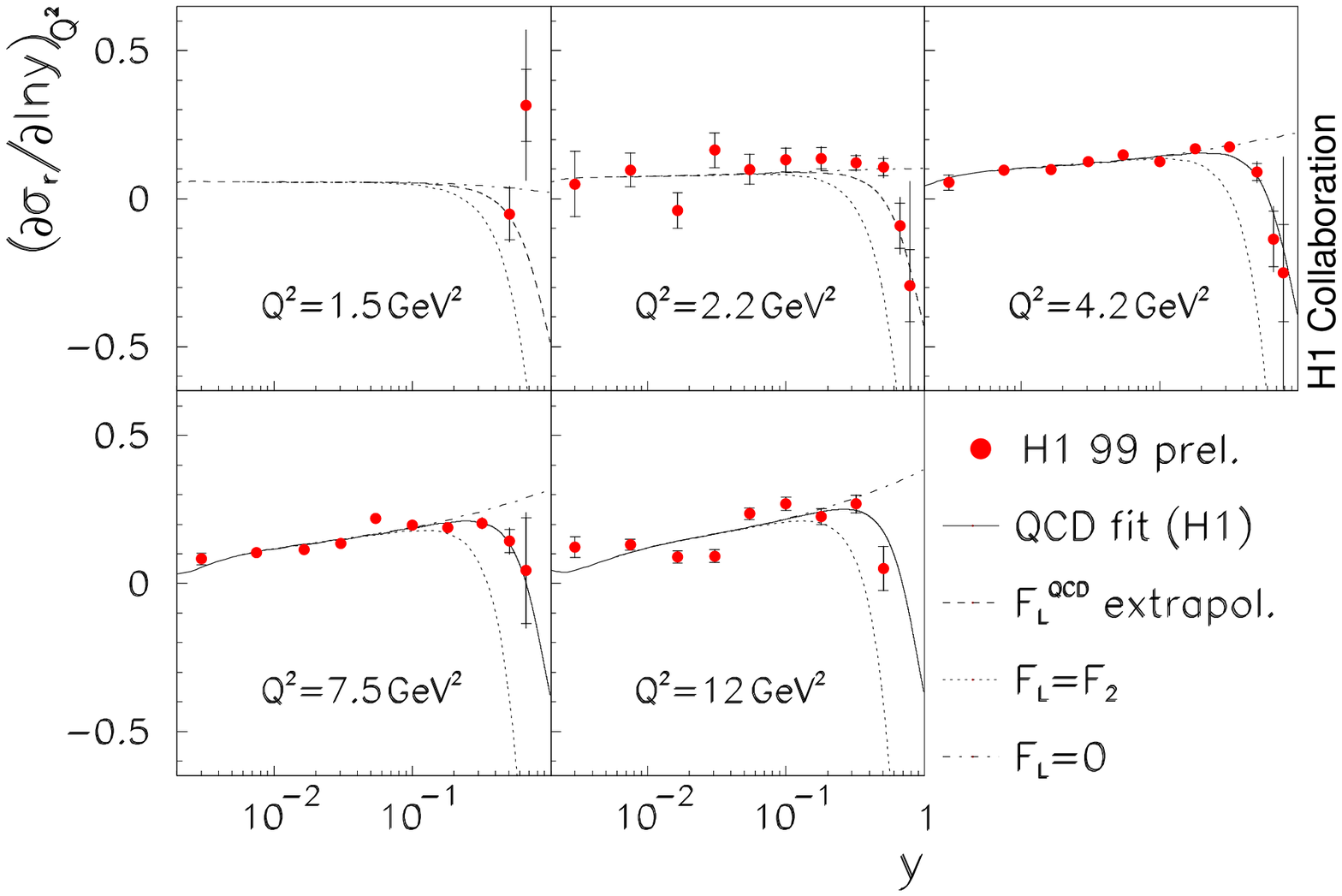,width=9cm}
\end{minipage}
\vspace*{-6cm}
\par\noindent
\hspace*{9.2cm}
\begin{minipage}[t]{4.7cm}
Figure 7: Measurement of $(\partial \sigma_{r} / 
\partial \ln y)_{Q^2}$ vs. $y$, in bins of $Q^2$. 
The curves represent the NLO QCD fit to the H1 
1996/97 data for $y <$ 0.35 and $Q^2 \ge$ 3.5 GeV$^2$, calculated for several
assumptions on $F_{L}$. 
\end{minipage}
\end{minipage}
\par\noindent
\begin{minipage}[t]{14cm}
\begin{minipage}[t]{13cm}
\begin{minipage}[t]{13cm}
\epsfig{file=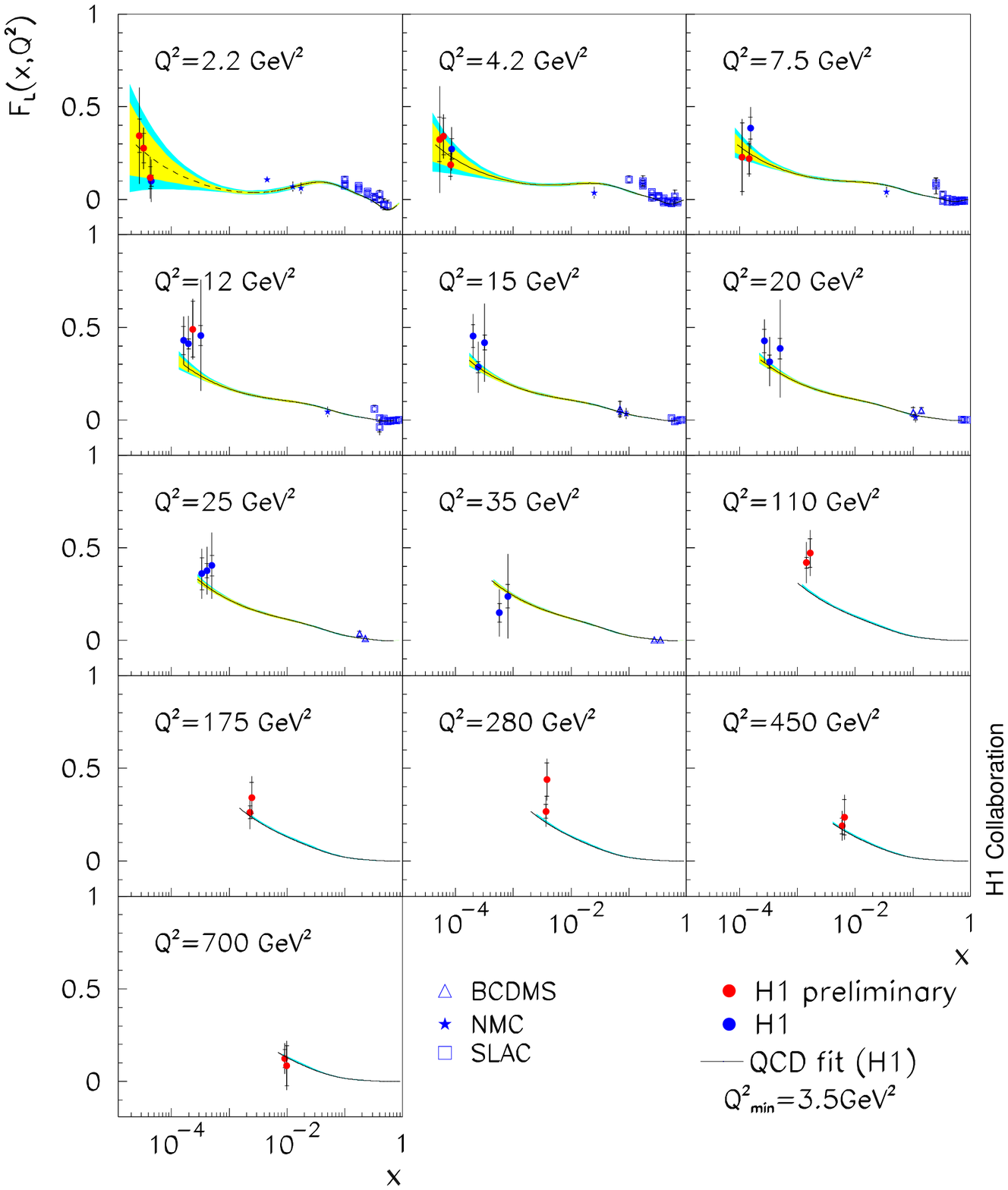,width=13cm}
\end{minipage}
\par\noindent
\hspace*{0.5cm}
\begin{minipage}[t]{12cm}
Figure 8: Longitudinal Structure Function $F_L$ vs. $x$ and in bins of $Q^2$,
obtained by H1 and fixed target experiments. Error bands are due to 
experimental (inner) and model (outer) uncertainties using the H1 NLO QCD
fit to the H1 1996/97 data for $y <$ 0.35 and $Q^2 \ge$ 3.5 GeV$^2$.
\end{minipage}
\end{minipage}
\end{minipage}
\medskip
\par\noindent
Fig.~8 gives an overview of the current H1 data on $F_L(x,Q^2)$ in the
$Q^2$ range $2.2 - 700$ GeV$^2$\cite{H100181,H1FLBudapest01}. 
The data extend the
knowledge of $F_L$ into the region of low~$x$,
much beyond the range of the fixed target  
experiments. The increase of $F_{L}(x,Q^2)$ towards low~$x$ 
is consistent with
the NLO QCD calculation, 
reflecting the rise of the gluon density
in this region. The values of $F_{L}(x,Q^2)$ are thus severely constrained
by the present data, unless there are deviations from the assumed 
extrapolation of $F_2$ into the region of large $y$ corresponding to the 
smallest~$x$.
\par\noindent
$F_L$ has also been calculated in the ZEUS NLO QCD fit\cite{ZEUSNLOfit2001};
the calculation is consistent with the H1 data. 

\section{Inclusive Jets and Dijets in DIS}

\bigskip
\bigskip
\par\noindent
\begin{minipage}[t]{14cm}
\begin{minipage}[t]{4cm}
\begin{minipage}[t]{4cm}
\epsfig{file=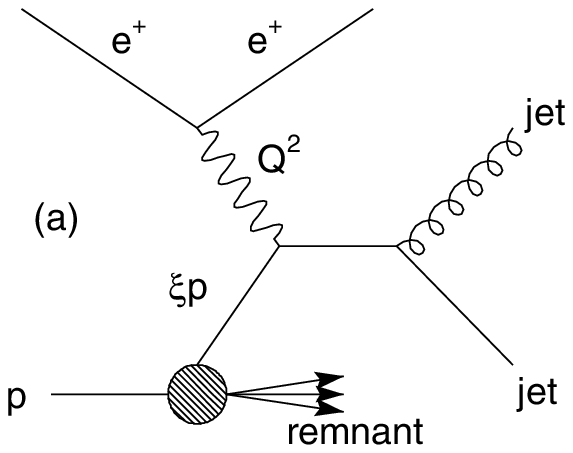,width=4cm}
\end{minipage}
\medskip
\par\noindent
\begin{minipage}[t]{4cm}
\epsfig{file=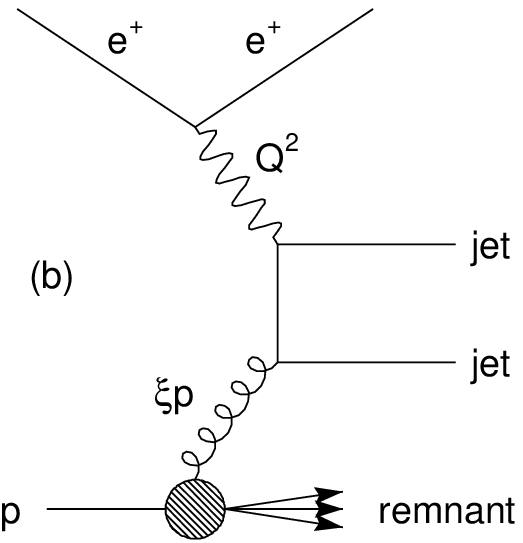,width=4cm}
\end{minipage}
\medskip
\par\noindent
\hspace*{0.3cm}
\begin{minipage}[t]{3.5cm}
Figure 9: Diagrams for QCD Compton and Boson Gluon Fusion processes.
\end{minipage}
\end{minipage}
 \vspace*{-11.1cm}
\par\noindent
\hspace*{4.3cm}
\begin{minipage}[t]{9.5cm}
The fully inclusive DIS process, shown in Fig.~1, is of zero order in 
$\alpha_s$. There is direct sensitivity only to the quark PDF's, and the
determination of $\alpha_s$ and gluon density is indirect, via the 
scaling violations of $F_2$, measured over a large range of $x$ and $Q^2$.
\par\noindent
The DIS processes shown in Figs.~9a and b, QCD Compton (QCDC) scattering and 
Boson Gluon Fusion (BGF), are of order $\alpha_s$ and due to the latter 
contribution there is direct 
sensitivity to the gluon density of the proton. At high enough energy, the
involved final state partons manifest as jets of hadrons. 
The multi-jet final state can be characterized by the jet mass, $M_{jj}$, and
the variable   
$\xi = x_{Bj} ( 1 + M_{jj}^2 / Q^2 )$. The dijet mass $M_{jj}$ gives the
CM energy of the boson-parton reaction and the fractional momentum $x$ 
(the longitudinal momentum fraction of the proton carried by the parton 
specified by the PDFs)
is
given by $\xi$, which is much larger than $x_{Bj}$ if $M_{jj}$ is large.
\par\noindent
Two hard 
scales enter in these diagrams, $Q^2$ and $E_{T,jet}$. Studies of the 
dynamics of multi-jet production are usually performed in the Breit frame,
where the virtual boson interacts head-on with the proton\cite{Breitframe}.
In lowest order, ${\cal O}(\alpha_s^0)$, \, the parton from 
\end{minipage}
\end{minipage}
\smallskip
\par\noindent
the proton is
backscattered into opposite direction,
and no transverse energy is produced.
Thus, appearance of jets with large $E_T$ can only be explained by hard
QCD processes of at least ${\cal O}(\alpha_s)$, and $E_{T,jet}$ is then the 
physical scale at which e.g. hard gluon radiation is resolved. 
Experimentally, jets are 
usually identified using a $k_T$ algorithm\cite{ktalgo}.
\par\noindent
The H1 and ZEUS collaborations have recently presented high statistics 
studies of inclusive jet and dijet production in 
NC DIS\cite{H1jet2000,ZEUSincljet2001,ZEUSdijet2001a,ZEUSdijet2001b}.
The data span the kinematic range $5-10<Q^2<10-15\cdot 10^3$~GeV$^2$ and 
$5-7<E_{T,jet}<60$~GeV.
NLO QCD calculations in general describe the data very well, over almost
the whole  
range of $Q^2$ and $E_{T,jet}$. This is exemplified in Fig. 10, 
showing the H1 inclusive jet cross section as function of $E_{T,jet}$ and
$Q^2$, and in Fig.~11, showing 
the ZEUS dijet cross section as function of $Q^2$.
The largest uncertainty in the QCD calculations stems from the uncertainty
in the choice of renormalization scale $\mu_r$, which is taken either as 
$Q$ or as mean jet transverse energy, $\overline{E}_{T,jet}$. The uncertainty
is largest for low $Q^2$ values (Fig.~11b), and cuts are made at
150 (H1) and 470 (ZEUS) GeV$^2$ in the QCD analyses of the data.
At large values of $Q^2$ the experimental uncertainties are also smaller,
and this is true as well for the hadronic (parton-to-hadron) corrections
to the NLO calculations, as shown in Figs. 10 and 11c.
\par\noindent
Once the good agreement of the QCD calculation with the jet data has been 
established, one can proceed to perform a QCD analysis, determining 
$\alpha_s$ and the gluon density. The analysis strategy is as follows:
\begin{enumerate}
\item
Fit the jet data to the NLO QCD predictions, using PDFs obtained from
global, external fits\cite{PDFjet,BOTJE99}. 
With the PDFs externally fixed, there is only one
free parameter, namely $\alpha_s$. Note that the PDFs themselves depend 
implicitly on 
$\alpha_s$, and that this complication was properly taken into account
in the fits.
\item
After establishing that the fitted $\alpha_s$ value agrees well with 
other, external measurements, $\alpha_s$ is fixed (e.g. to the world 
average) and the jet data are then fitted in order
to extract the PDFs, in particular the gluon density of the proton.
\item
Finally, a global, simultaneous fit of both $\alpha_s$ and
the PDFs can be performed. 
This would be a more independent test of QCD with the data. 
However, 
although in this global fit the quark PDFs emerge as identical to those
resulting from the fit in step (2), 
the simultaneous fit 
fails to produce meaningful results for $\alpha_s$ and the gluon density.
The reason   
is
\vspace*{0.3cm} 
\par\noindent
\hspace*{-0.7cm}
\begin{minipage}[t]{14cm}
\begin{minipage}[t]{7cm}
\begin{minipage}[t]{7cm}
\epsfig{file=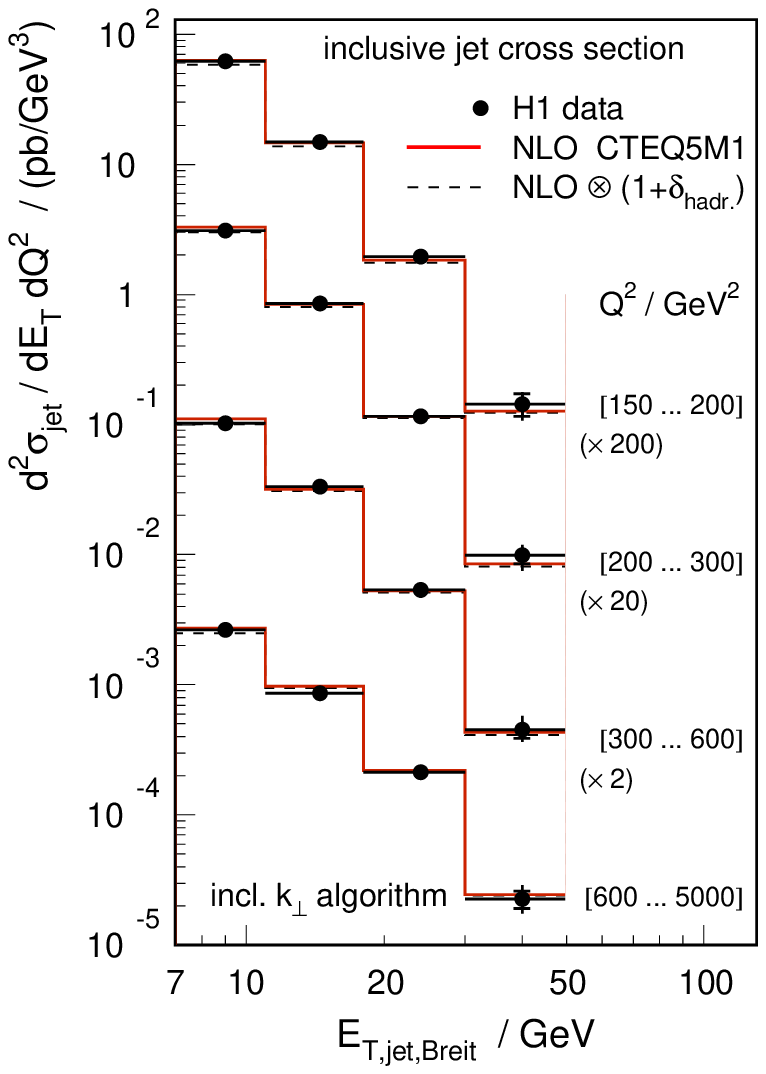,width=7cm}
\end{minipage}
\par\noindent
\hspace*{0.5cm}
\begin{minipage}[t]{6cm}
Figure 10: Inclusive jet cross section vs. $E_{T,jet}$, in several intervals
of $Q^2$. Also shown are NLO QCD calculations, with and without hadronic
corrections.
\end{minipage}
\end{minipage}
\vspace*{-12.2cm}
\par\noindent
\hspace*{7cm}
\begin{minipage}[t]{7cm}
\begin{minipage}[t]{7cm}
\hspace*{-0.4cm}
\epsfig{file=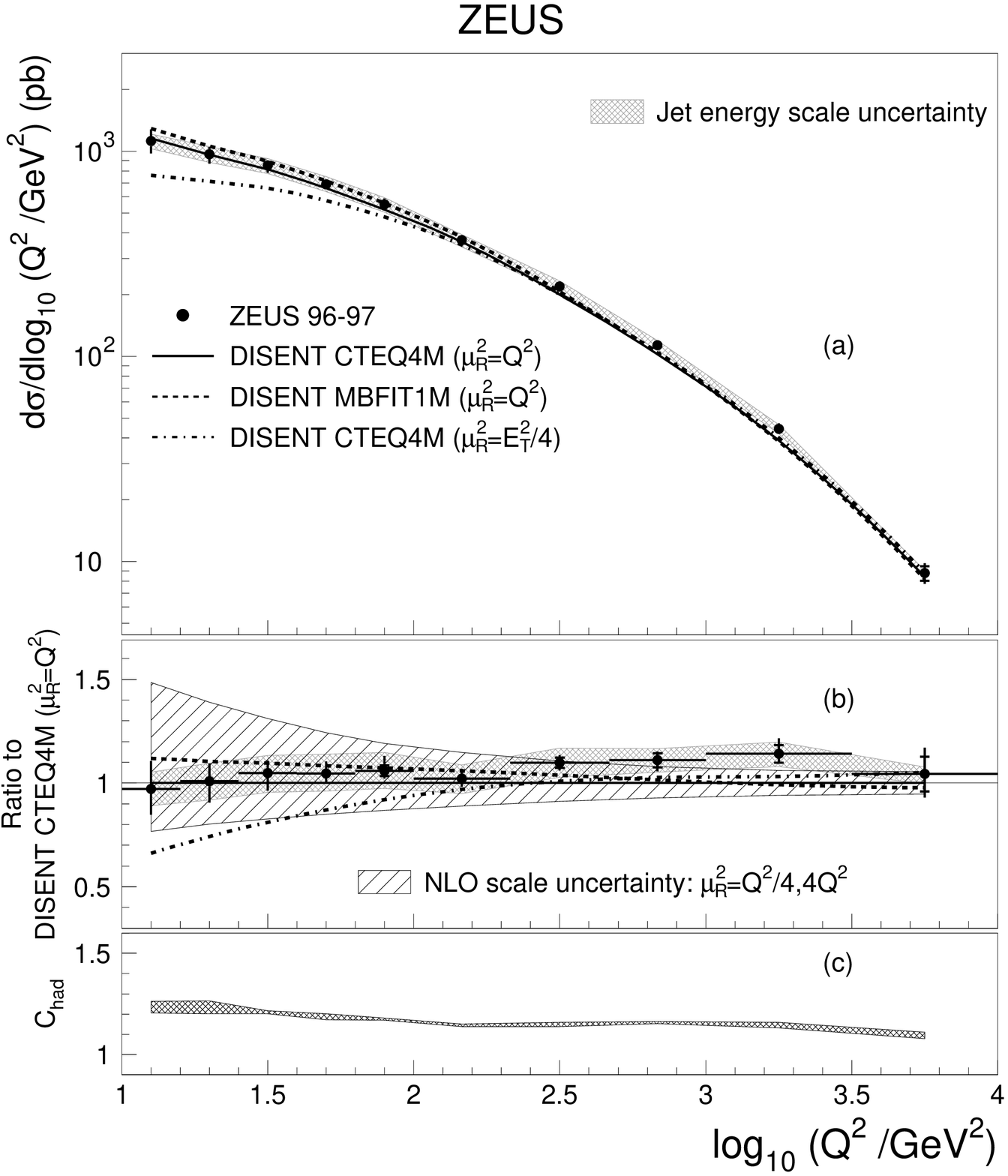,width=7cm}
\end{minipage}
\par\noindent
\hspace*{0.5cm}
\begin{minipage}[t]{6cm}
Figure 11: Dijet cross section vs. $Q^2$. 
The shaded band represents the calorimeter energy scale uncertainty. The
curves show NLO QCD calculations for several choices of $\mu_r$ and proton
PDF sets. b) Ratio data/theory showing in addition
the effect of the theoretical scale uncertainty. \newline
c) Parton-to-hadron 
correction.
\end{minipage}
\end{minipage}
\end{minipage}
\par\noindent
the strong anti-correlation between $\alpha_s$ and the gluon density,
which can be understood from the fact that in the phase space region 
considered, jet production is dominated by the gluon contribution, which
enters in the cross section as the product $\alpha_s\cdot x g(x)$.
\end{enumerate}
\begin{minipage}[t]{14cm}
\begin{minipage}[t]{14cm}
\begin{minipage}[t]{7.5cm}
\epsfig{file=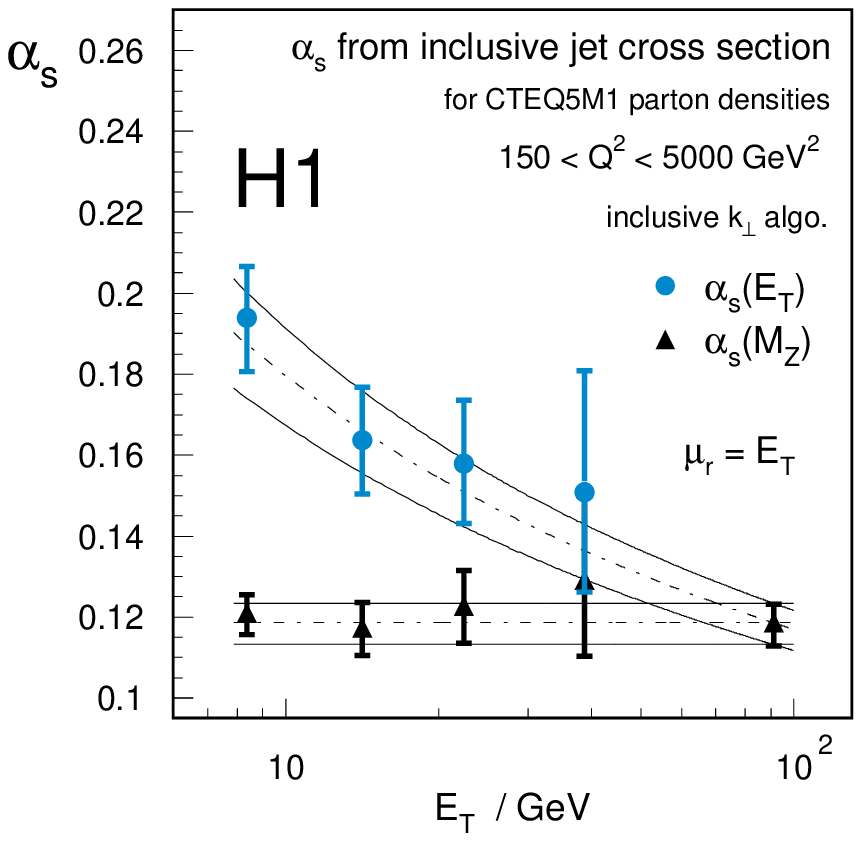,width=7.5cm}
\end{minipage}
\vspace*{-8.6cm}
\par\noindent
\hspace*{7.5cm}
\begin{minipage}[t]{7cm}
\epsfig{file=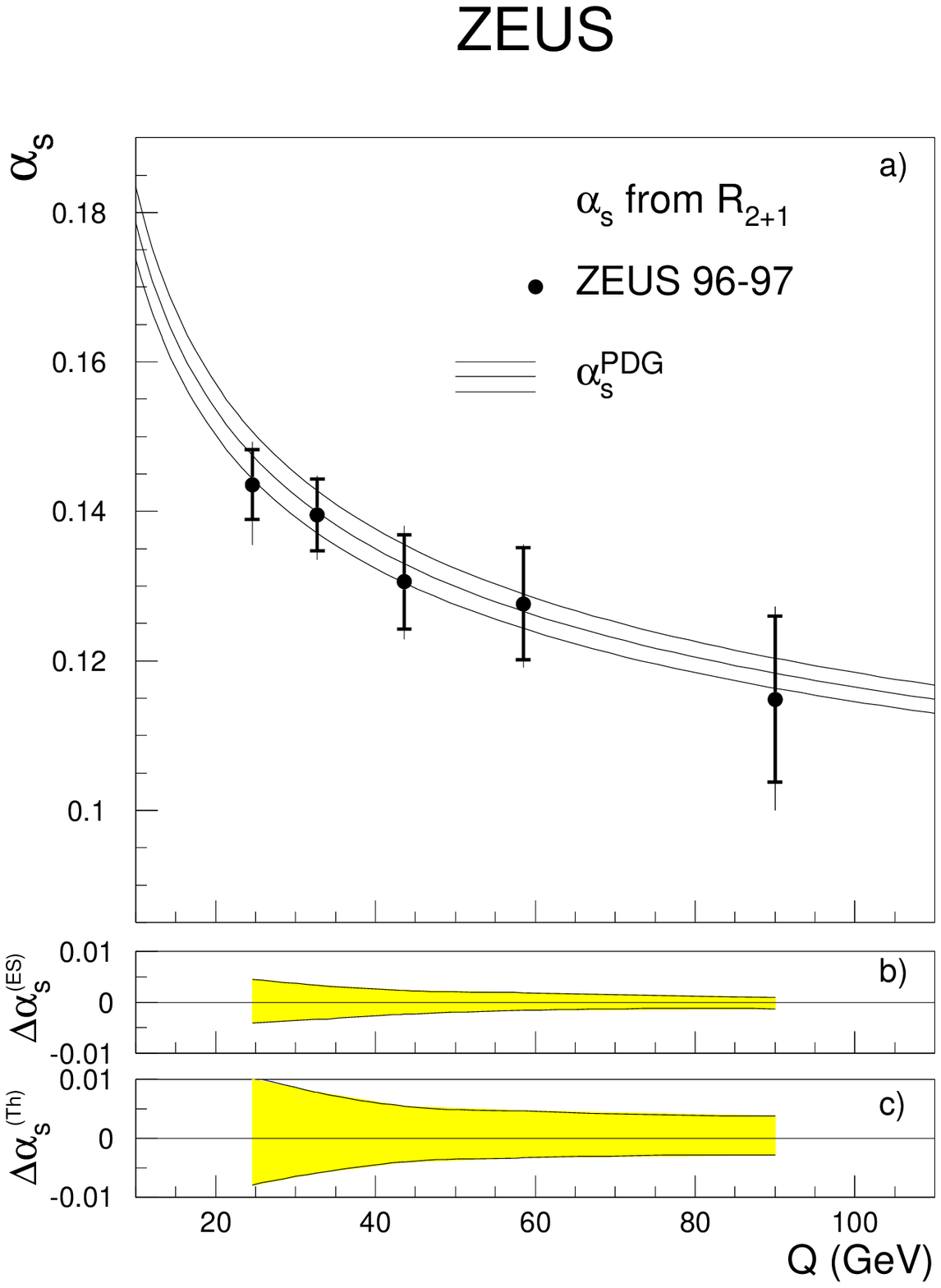,width=7cm}
\end{minipage}
\end{minipage}
\vspace*{-0.5cm}
\par\noindent
\hspace*{0.5cm}
\begin{minipage}[t]{13cm}
Figure 12: $\alpha_s$ running in 2 scales, $E_{T,jet}$ and $Q$. 
Uncertainties due to calorimeter energy scale and to theory are 
indicated in the ZEUS plot. Upper curves show the  RGE prediction.
\end{minipage}
\vspace*{0.5cm}
\par\noindent
\begin{minipage}[t]{6.5cm}
\begin{minipage}[t]{6.5cm}
\epsfig{file=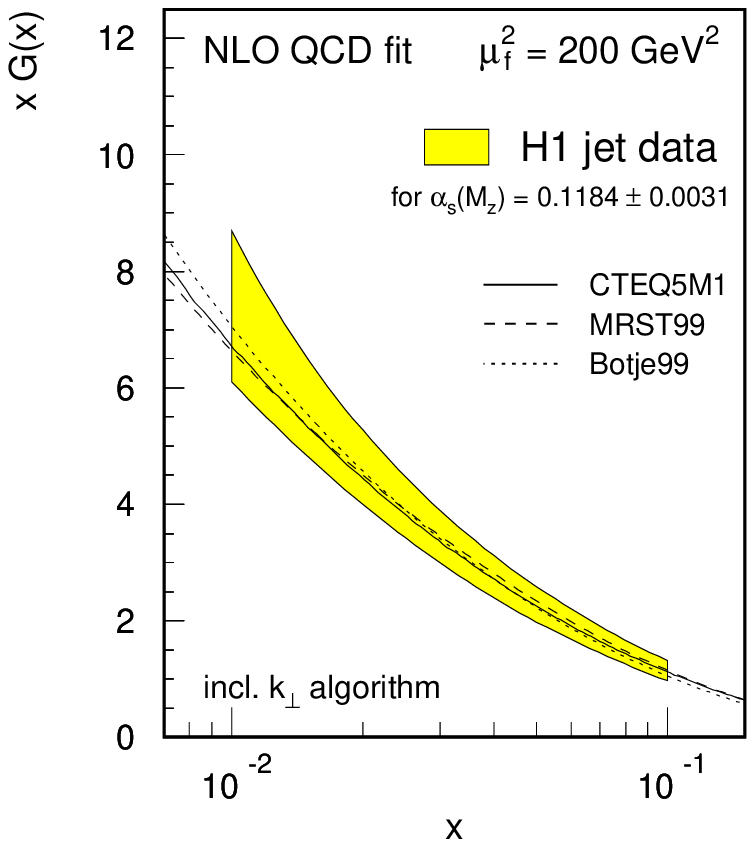,width=6.5cm}
\end{minipage}
\par\noindent
\hspace*{0.3cm}
\begin{minipage}[t]{6cm}
Figure 13: Gluon density in the proton, determined in a combined QCD fit 
using inclusive DIS , inclusive jet and dijet cross sections. The error
band includes all experimental and theoretical uncertainties, also 
that of $\alpha_s(M_Z)$.
\end{minipage}
\end{minipage}
\vspace*{-11cm}
\par\noindent
\hspace*{6.7cm}
\begin{minipage}[t]{7.2cm}
Both H1 and ZEUS have performed step~1 in this strategy, and the resulting 
$\alpha_s$ measurements are shown in Fig.~11. Note that 
the ``running'' of $\alpha_s$ is here clearly seen in both scales, $Q$ and 
$E_{T,jet}$, and the running is moreover seen within each single experiment.
The running is consistent with the renormalization group equation (RGE).
The comparison of the measurements with other HERA measurements, and with the
world average values is given in Fig.~15.
The fact that the $\alpha_s$ values obtained from the DIS jet
 data agree very well
with other measurements, in particular with measurements from processes which
do not involve hadrons in the initial state, like $e^+e^-$ annihilation to
hadrons, can be taken as proof of the validity of perturbative QCD at NLO in
DIS jet production.
\end{minipage}
\end{minipage}
\smallskip
\bigskip
\par\noindent
Step~2 in the strategy was performed by H1, and the gluon density extracted
from the data is shown in Fig.~13. The fit includes the combined 
cross sections of inclusive jet data, dijet data and inclusive DIS.
The result is in good agreement with recent global 
analyses\cite{PDFjet,BOTJE99,MRST99}. For a detailed comparison of this result
with the gluon densities obtained from the $F_2$ scaling violations, see
\cite{OSAKA992}.

\section{Jet Substructure}

\begin{minipage}[t]{14cm}
\begin{minipage}[t]{7cm}
\epsfig{file=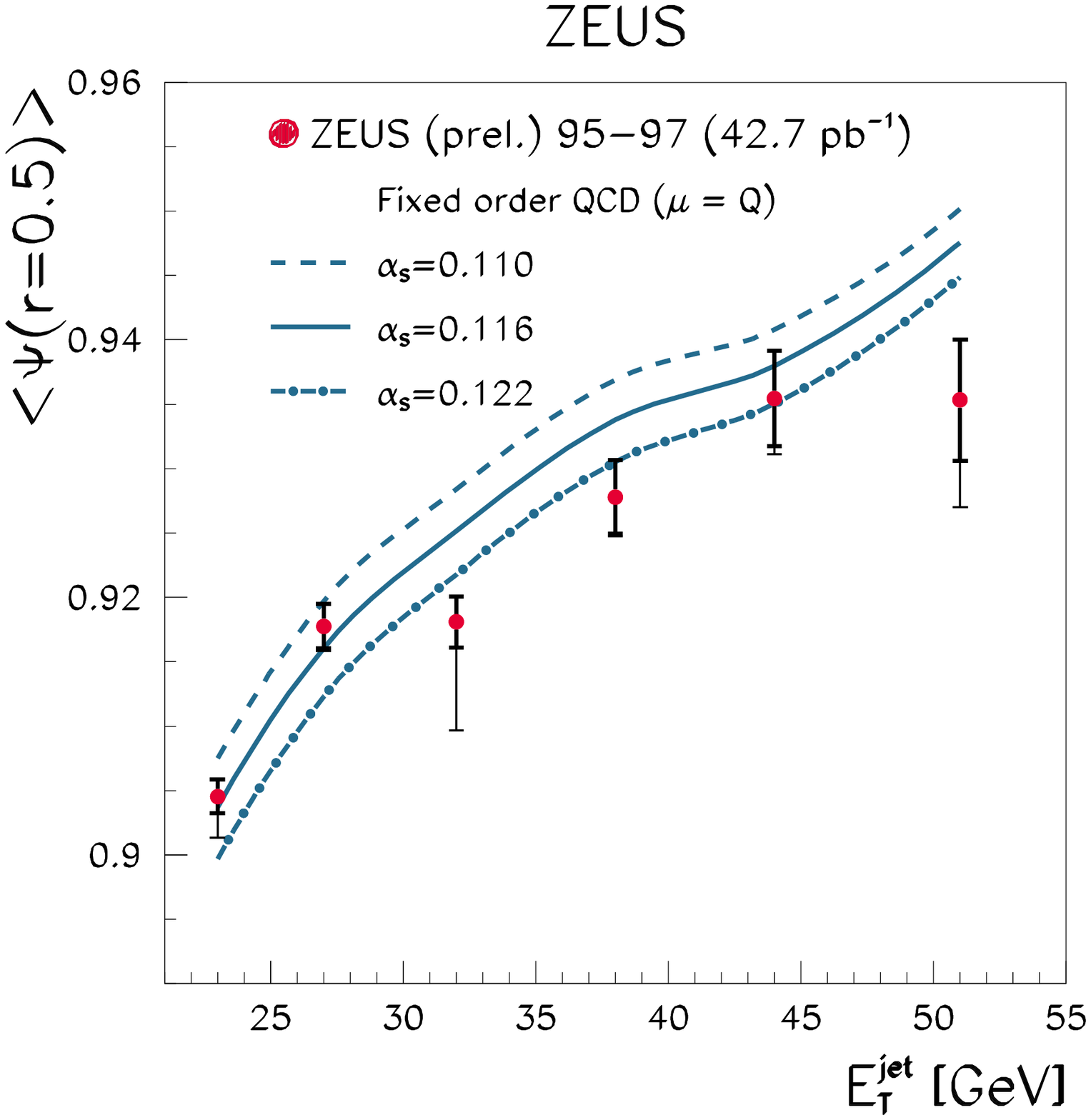,width=7cm}
\end{minipage}
\vspace*{-7.5cm}
\par\noindent
\hspace*{7cm}
\begin{minipage}[t]{7cm}
\epsfig{file=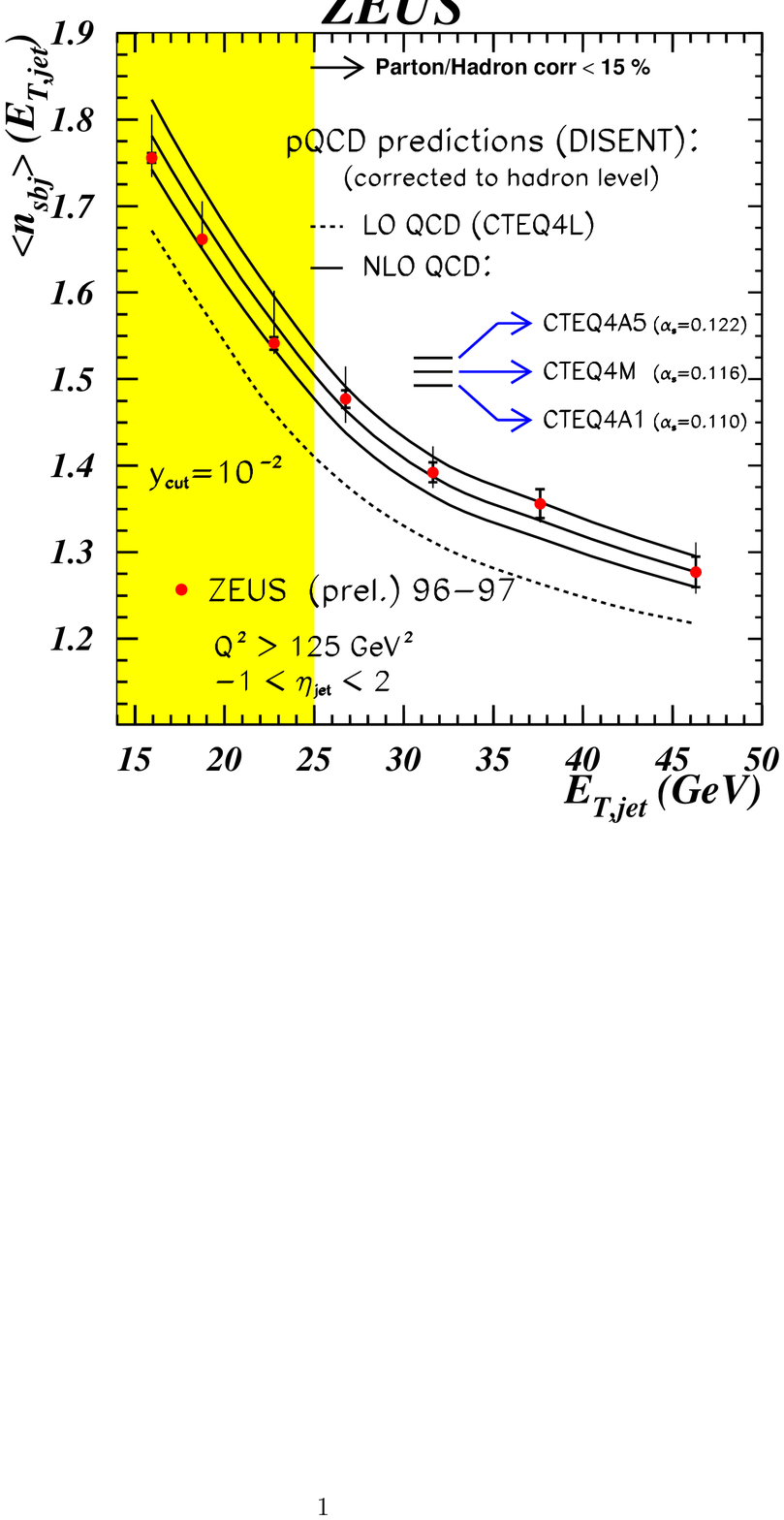,width=7cm}
\end{minipage}
\par\noindent
\hspace*{0.5cm}
\begin{minipage}[t]{13cm}
Figure 14: a) Integrated jet shape and b) mean subjet multiplicity as
functions of $E_{T,jet}$. NLO QCD calculations are shown for three 
different values of $\alpha_s$.
\end{minipage}
\end{minipage}
\bigskip
\par\noindent
At sufficiently high transverse jet energy, 
$E_{T,jet} \  ({\rm or}\ E_T^{jet})$, fragmentation 
effects become negligible and both the shape and other features of the
internal jet structure are expected to be calculable in perturbative QCD.
The ZEUS
collaboration has recently presented\cite{ZEUSsubjets} several results
based on studies of jet substructure, using the 
variables\cite{jetshape,subjetmult} ``integrated jet shape'' $\Psi(r)$
and ``mean subjet multiplicity'' $\langle n_{subjet}\rangle$:
\smallskip 
\par\noindent
\centerline{$
\Psi(r) = \frac{1}{N_{jet}}\sum_{1}^{N_{jet}}
\frac{E_T^{jet}(r)}{E_T^{jet}(r=R)}$ \ \ and \ \ 
$
\langle n_{subjet}\rangle = \frac{1}{N_{jet}}\sum_{1}^{N_{jet}}n_{subjet}.
$}
\smallskip
\par\noindent
The radius $r$ is defined in $\phi - \eta$ space, where the jet search  
is performed. The subjets within a given jet are found by repeating the jet
algorithm with smaller resolution scale. 
\par\noindent
Among the results presented in \cite{ZEUSsubjets} are
\begin{itemize}
\item
The average subjet multiplicity in jets in Charged Current (CC) and NC events
is found to be similar. Since the jets in CC and NC DIS are predominatly
quark initiated, the similarity in jet substructure indicates that the 
pattern of parton radiation within a quark jet is independent of the 
specific hard scattering process.
\item
In a dijet event sample, $c$-quark jets were tagged through the identification
of a $D^{\ast}$ meson. The internal structure of the charm induced jets is
found to be similar to that of the quark jets in NC DIS. Since the latter are
dominantly light quark initiated, one can conclude that the evolution of the
outgoing partons, which determines the internal structure of the jet, is 
independent of the hard subprocess from which the outgoing partons originate.
\item
Using the internal jet structure of the tagged, charm induced jets in the 
dijet event sample, and 
comparing with the internal jet structure of the total dijet sample, it was
possible to extract the internal jet structure of gluon jets. 
The prediction of
QCD, that gluon jets are broader, and contain more subjets, is nicely 
confirmed. 
\end{itemize}
\medskip
\par\noindent
\begin{minipage}[t]{14cm}
\begin{minipage}[t]{9cm}
\hspace*{-1.5cm}
\epsfig{file=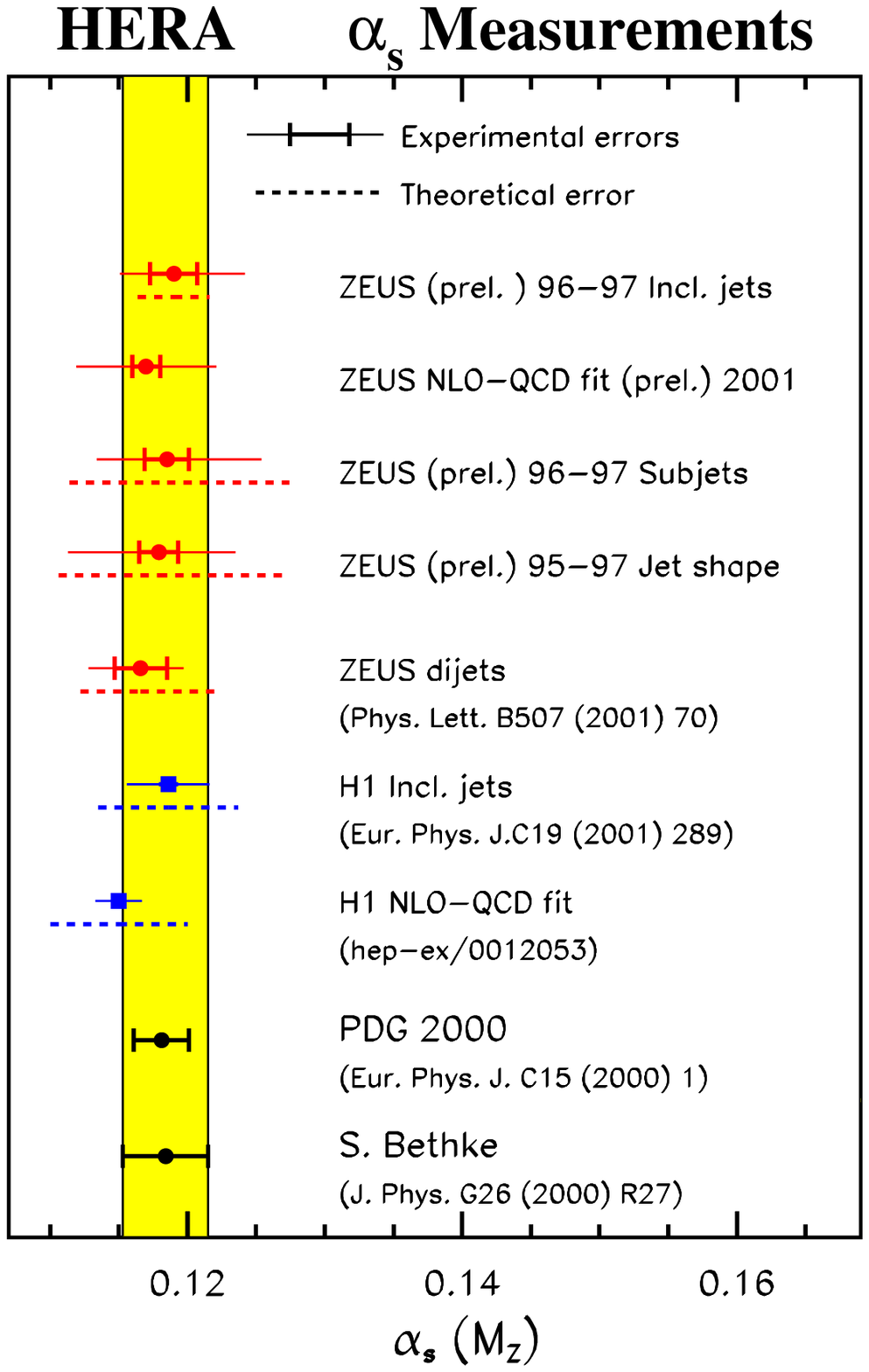,width=9cm}
\end{minipage}
\vspace*{-0.5cm}
\par\noindent
\begin{minipage}[t]{6cm}
Figure 15: Summary of $\alpha_s$ measurements at HERA (given at the 
$Z$ mass), using inclusive
DIS data and NLO QCD fits, inclusive jets and dijets, and jet shape and 
subjet multiplicity. Also shown are the world average values 
(PDG 2000, S. Bethke).
\end{minipage}
\vspace*{-13.5cm}
\par\noindent
\hspace*{6.7cm}
\begin{minipage}[t]{7cm}
The internal jet structure is sensitive to $\alpha_s$ beyond leading order. 
This sensitivity is demonstrated in Fig.~14 a and b, where the integrated
jet shape (for $r=0.5$) and mean subjet multiplicity is shown as functions of
$E_{T,jet}$. Note that the jets become narrower, and that the mean subjet 
multiplicity decreases as $E_{T,jet}$ increases. The NLO QCD calculation 
 describes both jetshape and mean subjet multiplicity well, but the
calculation also varies strongly with different values of
$\alpha_s$. It is thus clear that the jet substructure 
data can be used to determine 
$\alpha_s$. The method is similar to the one used in the QCD analysis of
the inclusive and dijet data described above. External sets of PDFs are
used, and the implicit dependence on $\alpha_s$ in these sets is properly
taken account of in the fitting.
\par\noindent
The results of the $\alpha_s$ determination using the jet substructure are
shown in Fig.~15, together with other $\alpha_s$ measurements from HERA.
These new measurements agree well with the others, 
which is proof of the consistency of the NLO QCD calculations 
also where the internal jet structure is concerned.
\end{minipage}
\end{minipage}
\medskip
\par\noindent 
A general remark about the measurements in 
Fig.~15 can be made: 
The experimental errors in these measurements
are small and comparable to the error on the
world averages\cite{PDG,Bethke00}. The total 
errors on the $\alpha_s$ measurements at HERA are in fact everywhere 
dominated by the theoretical uncertainty. As already noted above, this 
situation is expected to change when  NNLO calculations become available.

\section*{The Three-Jet Final State in DIS}

\begin{minipage}[t]{14cm}
\begin{minipage}[t]{7cm}
\begin{minipage}[t]{7cm}
\epsfig{file=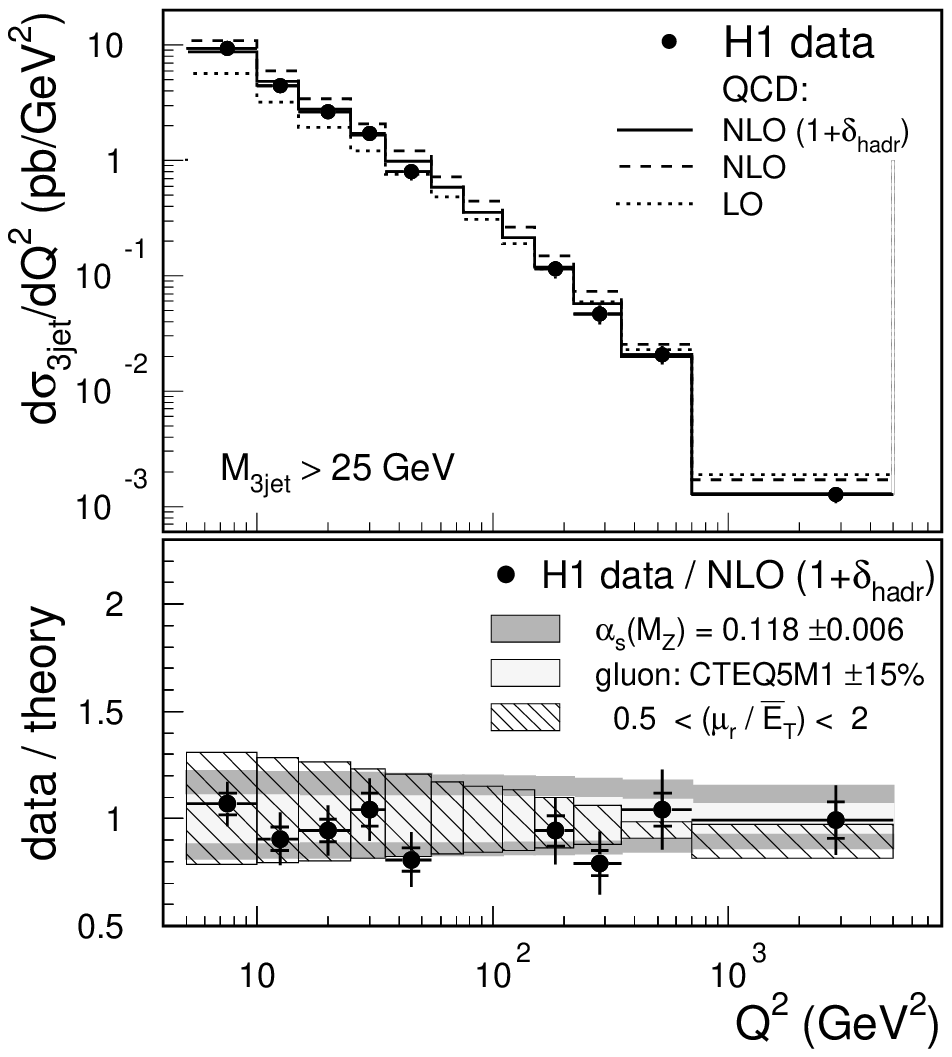,width=7cm}
\end{minipage}
\par\noindent
\begin{minipage}[t]{7cm}
\epsfig{file=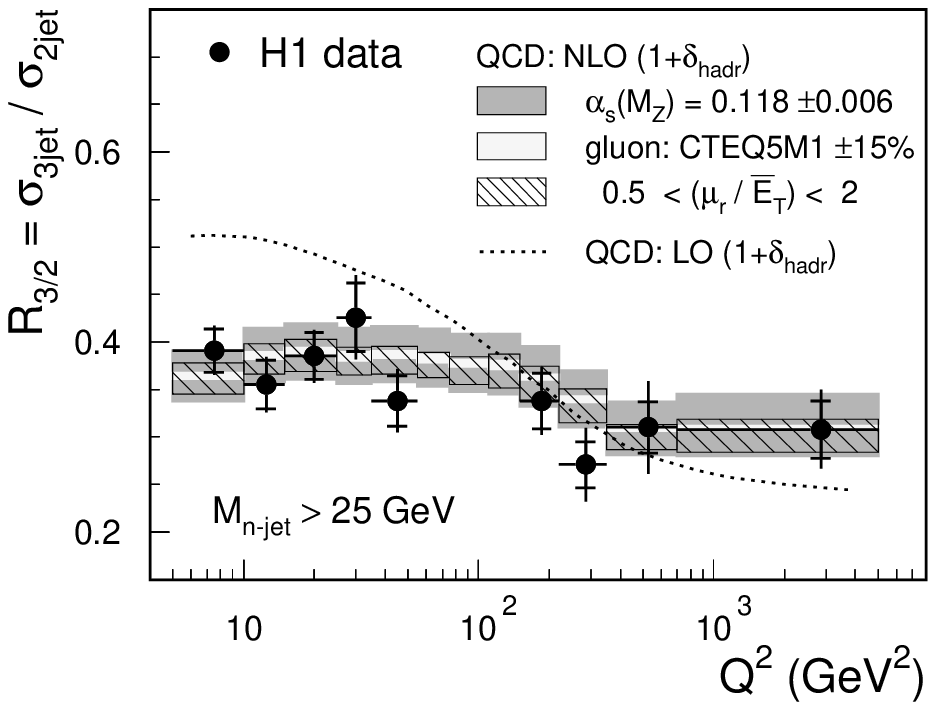,width=7cm}
\end{minipage}
\par\noindent
\hspace*{0.5cm}
\begin{minipage}[t]{6cm}
Figure 16: a) Inclusive three-jet cross section vs. $Q^2$. Also shown is the 
ratio of data to theoretical prediction.
b) Ratio of inclusive three-jet and dijet cross sections vs. $Q^2$.
LO and NLO QCD calculations are also shown. including the effects of variation 
in the latter 
of $\alpha_s(M_Z)$, renormalization scale $\mu_r$ and proton gluon density. 
\end{minipage}
\end{minipage}
\vspace*{-18.2cm}
\par\noindent
\hspace*{7.2cm}
\begin{minipage}[t]{6.7cm}
While the inclusive jet and dijet cross sections are directly sensitive to
QCD effects of order ${\cal O}(\alpha_s)$, the three-jet cross section in DIS
is already proportional to $\alpha_s^2$ in leading order in perturbative QCD.
This higher sensitivity to $\alpha_s$ and 
the greater number of degrees of freedom
of the three-jet final state  allow the QCD predictions to be tested in 
more detail in three-jet production. The H1 collaboration has recently 
presented\cite{H13jet} a study of DIS three-jet events,
covering the kinematic range
$5<Q^2<5000$ GeV$^2$ and three-jet masses $25<M_{3jet}<140$ GeV. The $Q^2$ 
dependence of the cross section is shown in Fig.~16a, together with QCD 
LO and NLO calculations, the latter with and without hadronic corrections. 
The NLO calculation, which is due to the recently available 
program NLOJET\cite{NLOJET}, describes the data well over the whole kinematic
range, as is also seen in the ratio of data to theory. In the latter plot are
also shown the theory uncertainty due to variation of the 
gluon density, the renormalization 
scale $\mu_r=\overline{E}_T$  and $\alpha_s$. As seen,
at large $Q^2>50$ GeV$^2$  the $\alpha_s$ variation gives the largest
uncertainty.
\par\noindent
Since both dijet and three-jet production is dominated by gluon induced
processes,
the uncertainty of the gluon density can be minimized by taking the ratio
$R_{3/2}$ of three-jet 
\end{minipage}
\end{minipage}
\smallskip
\par\noindent
and dijet cross sections, at the same values of $x$ and $M_{n-jet}$.
It can be shown, using the QCD calculations, that dijet and three-jet
production involves the same gluon fraction, at similar $Q^2$ values.
Furthermore, many experimental and systematic errors cancel in the ratio. 
As is evident in
Fig.~16b, $R_{3/2}$, which is directly proportional to $\alpha_s$,
 is experimentally measured and 
theoretically calculated with small uncertainties over the whole $Q^2$ range. 
Thus, given better  
statistics, a very sensitive test of QCD will be possible with the 
three-jet data, including a 
precision measurement of $\alpha_s$.
\par\noindent
\begin{minipage}[t]{14cm}
\begin{minipage}[t]{4.5cm}
\epsfig{file=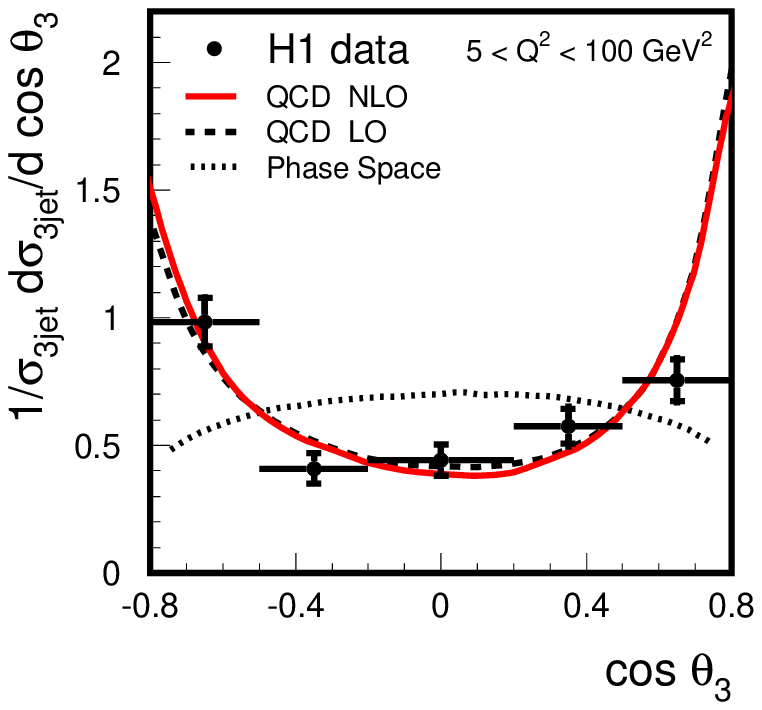,width=4.5cm,height=4.5cm}
\end{minipage}
\vspace*{-4.5cm}
\par\noindent
\hspace*{4.4cm}
\begin{minipage}[t]{4.5cm}
\epsfig{file=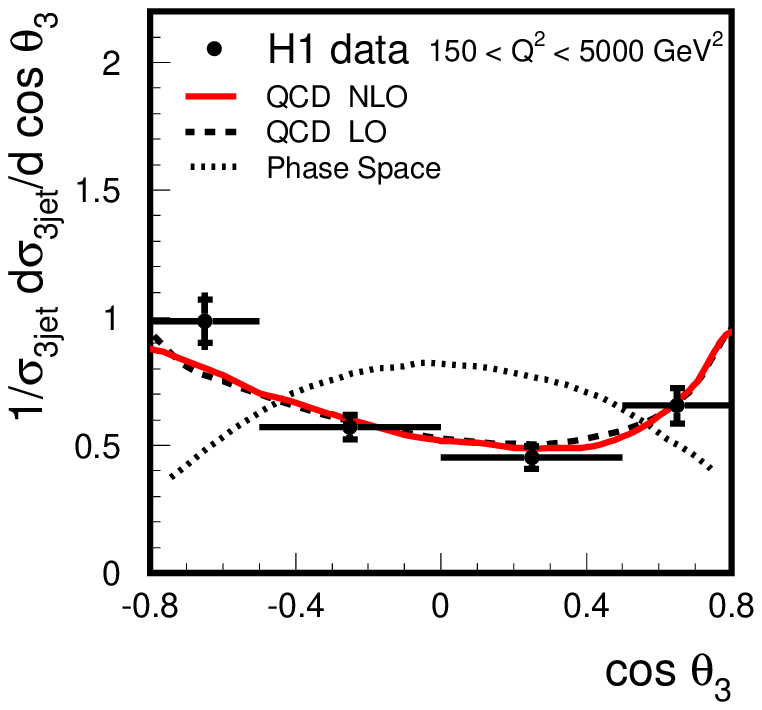,width=4.5cm,height=4.5cm}
\end{minipage}
\vspace*{0.1cm}
\par\noindent
\begin{minipage}[t]{4.5cm}
\epsfig{file=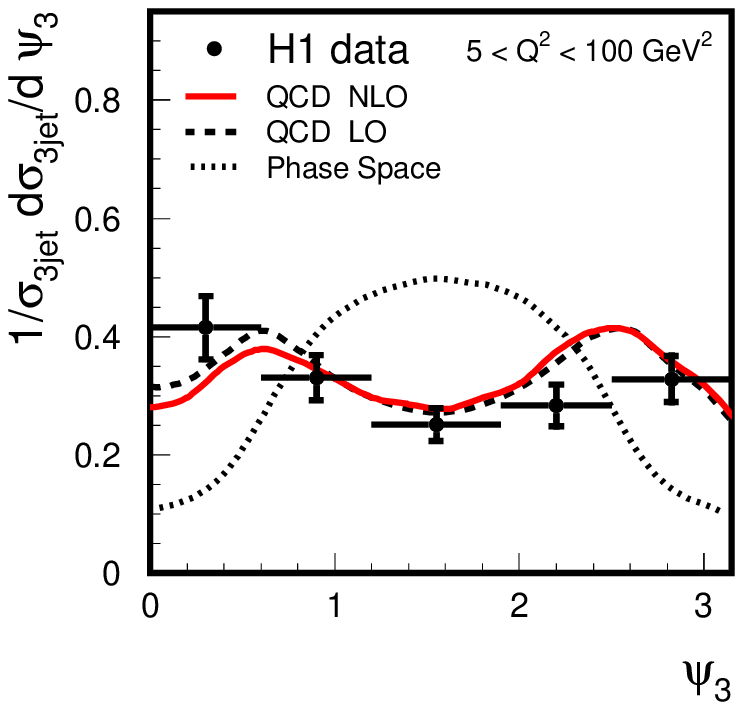,width=4.5cm,height=4.5cm}
\end{minipage}
\vspace*{-4.5cm}
\par\noindent
\hspace*{4.4cm}
\begin{minipage}[t]{4.5cm}
\epsfig{file=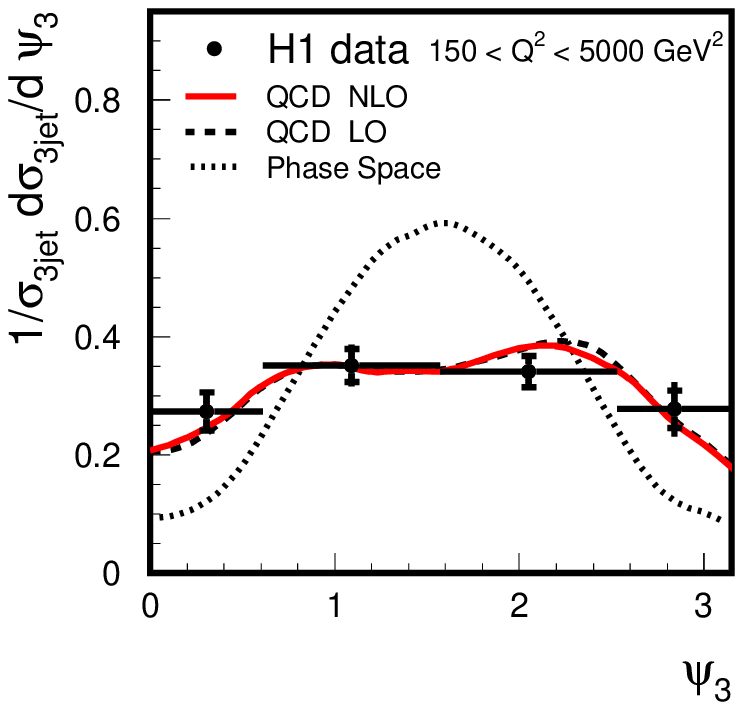,width=4.5cm,height=4.5cm}
\end{minipage}
\par\noindent
\hspace*{0.4cm}
\begin{minipage}[t]{8.3cm}
Figure 17: Normalized distributions of $\cos \theta_3$ 
and angle~$\psi_3$ in the
three-jet CMS at low $Q^2$ and high $Q^2$. Solid and dashed curves show
NLO and LO QCD calculations, dotted curves show a three-jet phase space
model.
\end{minipage}
\vspace*{-11.8cm}
\par\noindent
\hspace*{9.2cm} 
\begin{minipage}[t]{4.6cm}
The topology of the three-jet final state offers a test of QCD through
the angular distributions of $\theta_3$ and $\psi_3$, where 
 $\theta_3$ is the angle between the highest energy jet and the proton
beam, and $\psi_3$ is the angle between the two planes formed by the
highest energy jet and the proton beam, and by the three jets, respectively.
These angles are shown in Fig.~17, for two $Q^2$ ranges. Both LO and NLO QCD
calculations describe these normalized distributions well, while a phase
space model fails the description. The distributions show that the jets tend
to be aligned with either the photon or
\end{minipage}
\end{minipage}
\smallskip
\par\noindent
the proton, i.e. the Bremsstrahlung nature
of the process (coherence property of QCD) is confirmed. Similar distributions
were previously 
also observed by the ZEUS collaboration
 in three-jet photoproduction\cite{ZEUSgammap3jet}.

\section*{Conclusions}

The DIS data are well described by NLO QCD and the DGLAP evolution, over
a large range of $Q^2$ and Bjorken-$x$. This is true for the inclusive
DIS cross section, with the structure functions $F_2$ and $F_L$, as well
as for the inclusive jet production and the exclusive dijet and three-jet
production and for the internal structure of jets. 
\par\noindent
In many areas the data have reached a high experimental precision, and 
progress in the tests of perturbative QCD depends crucially on 
further progress in the theory, where the NNLO calculations for the DIS
processes are awaited. This is particularly true for the precision 
determination of $\alpha_s$ at HERA, using DIS data, where the experimental
precision is already at level with the error on the world average, and
is expected to improve even further. 
\par\noindent 
The running of $\alpha_s$ in accordance with the RGE is seen within 
each single experiment, and in both hard scales, $Q$ and $E_{T,jet}$.
\par\noindent
Some areas, like jets at highest $Q^2$ and $E_{T,jet}$, and the 
three-jet 
final state, are still statistically
limited. The HERA II running period will bring a huge improvement
of statistics in
the coming years.  However, in the analyses described in this report, 
the experiments have in many cases so far only
used the data taken in 1995-97, which constitute only about 1/3 of the
total HERA~I data. The remaining 2/3 of the HERA~I data, from the years
1998-2000, is currently being 
recalibrated and reprocessed, and will soon be available. 

\section*{Acknowledgments}

It is a pleasure to thank the organizers for the warm and joyful 
atmosphere in a most interesting and remarkably well prepared conference.
I also wish to thank my
colleagues in H1 and ZEUS, for providing the data and results
presented in this report and for all their help given to me.

\end{document}